\documentclass[11pt, a4paper]{article} 

\pdfoutput=1

\usepackage{jcappub}
\usepackage{graphicx}   
\usepackage{subfigure}
\usepackage{hyperref}
\usepackage{float}
\usepackage{color}
\usepackage{amsmath,amsfonts,amsthm} 
\usepackage{slashed}

\usepackage{epsfig}

\begin{document}

\title{The role of fluctuation-dissipation dynamics in setting initial conditions for inflation}

\author[a]{Mar Bastero-Gil}    \emailAdd{mbg@ugr.es}
\affiliation[a]{Departamento de F\'{\i}sica Te\'orica y del Cosmos, 
Universidad de Granada, Granada-18071, Spain}

\author[b]{Arjun Berera}    \emailAdd{ab@ph.ed.ac.uk}
\affiliation[b]{School of Physics and Astronomy, University of Edinburgh, Edinburgh, 
EH9 3FD, United Kingdom}

\author[c]{Robert Brandenberger} \emailAdd{rhb@physics.mcgill.ca}
\affiliation[c]{Physics Department, McGill University, Montreal, QC, H3A 2T8, Canada}

\author[d]{Ian G. Moss} \emailAdd{ian.moss@ncl.ac.uk} 
\affiliation[d]{School of Mathematics, Statistics and Physics, Newcastle University,
Newcastle upon Tyne, NE1 7RU, United Kingdom}

\author[e, c]{Rudnei O. Ramos} \emailAdd{rudnei@uerj.br}
\affiliation[e]{Departamento de F\'{\i}sica Te\'orica, Universidade do
Estado do Rio de Janeiro, 20550-013 Rio de Janeiro, RJ, Brazil}

\author[(f)]{Jo\~{a}o G.~Rosa} \emailAdd{joao.rosa@ua.pt}
\affiliation[f]{Departamento de F\'{\i}sica da Universidade de Aveiro and CIDMA,  
Campus de Santiago, 3810-183 Aveiro, Portugal}


\arxivnumber{1612.04726}

\abstract{

We study the problem of initial conditions for slow-roll inflation
along a plateau-like scalar potential within the framework of
fluctuation-dissipation dynamics. We consider, in particular, that
inflation was preceded by a radiation-dominated epoch where the
inflaton is coupled to light degrees of freedom and may reach a
near-equilibrium state. We show that the homogeneous field component
can be sufficiently localized at the origin to trigger a period of
slow-roll if the interactions between the inflaton and the thermal
degrees of freedom are sufficiently strong and argue that this does
not necessarily spoil the flatness of the potential at the quantum
level. We further conclude that the inflaton can still be held at the
origin after its potential begins to dominate the energy balance,
leading to a period of thermal inflation. This then suppresses the
effects of nonlinear interactions between the homogeneous and
inhomogeneous field modes that could prevent the former from entering
a slow-roll regime. Finally, we discuss the possibility of an early
period of chaotic inflation, at large field values, followed by a
first stage of reheating and subsequently by a second inflationary
epoch along the plateau about the origin. This scenario could prevent
an early overclosure of the Universe, at the same time yielding a low
tensor-to-scalar ratio in agreement with observations.
}

\keywords{inflation, initial conditions, fluctuation-dissipation dynamics}


\maketitle


\section{Introduction}

The inflationary scenario~\cite{Guth} is the current paradigm of early
Universe cosmology. It addresses a number of problems of Standard Big
Bang cosmology, provided the first mechanism based on causal physics
to generate the observed density fluctuations on cosmological
scales~\cite{Mukh}, and made predictions concerning the spectrum of
cosmic microwave background (CMB) anisotropies which were subsequently
successfully verified  by the measurements from both
WMAP~\cite{Hinshaw:2012aka} and Planck~\cite{Ade:2015lrj} satellites
(among these measurements relevant to inflation, one can mention the
near Gaussianity of the fluctuations, the acoustic oscillations in the
angular power spectrum of CMB anisotropies, and the small red tilt of
the spectrum of fluctuations)\footnote{Note, however, that there are
  alternative early universe scenarios such as the Ekpyrotic
  scenario~\cite{Ekp}, String Gas Cosmology~\cite{SGC} and the Matter
  Bounce model~\cite{FB,Wands}, which also lead to fluctuations that
  are consistent with the current data.}.

The inflationary scenario assumes that the patch of the universe we
observe underwent a period of almost exponential expansion during some
time interval at some very early time. This accelerated expansion is
driven by the potential energy of a slowly rolling scalar matter
field. A key question when discussing the success of inflation is the
{\it initial condition question}, namely how likely are initial
conditions which lead to a sufficiently large region to enter a period
of inflation. There have been claims in the literature that the
initial conditions to obtain inflation have to be severely fine
tuned~\cite{Penrose, Gibbons, Trodden, Ijjas}. Indeed, for {\it small
  field models} of inflation, the initial field velocity has to be
severely fine tuned in order to obtain inflation~\cite{Piran}.

On the other hand, for {\it large field models} of inflation the slow
roll trajectory of the scalar field is a local attractor in
initial conditions space. This was indicated by the initial studies
done in Ref.~\cite{Albrecht:1984qt},  worked out in more detail in
Ref.~\cite{Brandenberger:1990wu}, and generalized to the presence of
linear gravitational fluctuations in Ref.~\cite{Feldman} (see also
Ref.~\cite{RHBICrev} for a recent review). These works used somewhat
intuitive reasonings, but investigations using numerical general
relativity~\cite{Matzner} confirmed the basic picture, and recent
numerical work shown in the Refs.~\cite{East,Lim} contributed to an
improved characterization of the attractor basin of large field
inflation.

Such large field models tend, however, to predict a too large fraction
of tensor modes in the primordial perturbation spectrum, and are thus
in tension with the most recent CMB observations by the WMAP and
Planck satellites. Observational data seems to suggest, in fact,
scalar potentials with a plateau-like region for which the
tensor-to-scalar ratio is suppressed. This poses a significant
challenge for inflationary model-building, since a slow-roll solution
is not a phase space attractor for most potentials in
this class. This means that inflation can only be triggered if the
field is initially located in the inflationary plateau and if its
velocity is sufficiently small. In addition, this must be true on
super-Hubble scales, since nonlinear effects (both in the field and in
the metric) could prevent the start of inflation. This means, in
particular, that even when the conditions leading to inflation are
satisfied within a Hubble-sized patch, if the field value outside this
region does not meet these requirements, the field and space-time
dynamics may eventually make the field leave the slow-roll regime
everywhere. In {}Fourier space, this can alternatively be seen as due
to nonlinear mixing between super-Hubble and sub-Hubble modes in the
field and Einstein equations.

These problems are generic for scalar potentials where inflation
occurs only at a scale $V_0\ll M_P^4$ so that, to make matters worse,
inflation can only be triggered long after the Planck era. {}For
plateau-like potentials inflation thus seems to require finely-tuned
initial conditions for the field value, its velocity and degree of
homogeneity, and also for the initial spatial curvature of the
Universe following the Planck era. Some authors have even gone as far
as saying that the inflationary paradigm is in considerable risk of
failing~\cite{Ijjas} (see, however,
Refs.~\cite{Guth:2013sya,Linde:2014nna} for a different point of
view).

Historically, considerations of the early universe have either treated
the cosmological dynamics in the thermal limit, such as during the hot
Big Bang regime, or in a zero temperature state, such as during
inflation. The proposal of the warm inflation
paradigm~\cite{Berera:1995wh,Berera:1995ie,Berera:1996nv} showed that
the dynamics slightly perturbed away from the thermal limit can have
important consequences with robust consistency with data. In
particular, Refs.~\cite{Bartrum:2013fia,Bastero-Gil:2016qru} have
shown that going beyond the conventional setup for inflation and
including interactions between the inflation and other degrees of
freedom generically lowers the tensor-to-scalar ratio in monomial
models, rendering them into agreement with observations. These
interactions, which must always be present to ensure the transition
from inflation to a radiation-dominated Universe, may lead to
finite-temperature fluctuation-dissipation effects that can (depending
on the model building details) potentially modify the dynamics of the
inflaton field, at the level of both the homogeneous background
value and the small perturbations about the latter. These effects are
associated with particle production and, in the warm inflation
scenario, are themselves capable of sustaining a nearly-thermal bath
during the inflationary expansion. 

More generally, there are several periods during the evolution of
the early Universe where the Universe is slightly away from
equilibrium and fluctuation-dissipation dynamics could be present.
Such regimes may be essential for explaining important features
of the observed Universe. In particular it has been shown that
fluctuation-dissipation dynamics can play a relevant
role in cosmic phase transitions \cite{Bartrum:2014fla} and
baryogenesis \cite{BasteroGil:2011cx} (for other applications of the
effects of noise and dissipation in cosmology, see, for example,
Refs.~\cite{ng1, ng2, Berera:2009gy, Ramos:2013nsa, Bastero-Gil:2014jsa,
  Bastero-Gil:2014raa, Vicente:2015hga}).

The question that naturally arises from these previous studies is
whether fluctuation-dissipation effects can play any role in the
pre-inflationary era and, in particular, whether they offer any novel
insight on the problem of initial conditions, in particular for the
plateau-like potentials for which this problem is most severe as
discussed above. The presence of a friction term in the inflaton's
equation of motion has already been shown to have a substantial impact
in determining the minimum size of the initial inflationary
patch~\cite{Berera:2000xz}. In Ref.~\cite{Berera:2000xz}, it has been
argued that if before inflation strong dissipation (characterized by a
dissipation coefficient $\Upsilon$ compared to the Hubble expansion
rate $H$ satisfying $\Upsilon > H$) is present, then it will damp
fluctuations of modes with physical wavenumber $k < \Upsilon$.  This
means smoothness of the initial patch size need not be required at the
Hubble scale $1/H$ but over the much smaller scale $1/\Upsilon$.
{}For scales bigger than that, since dissipation will damp the modes,
it will prevent nonlinear dynamics from becoming significant. The
works in Ref.~\cite{Albrecht:1984qt} were amongst the first to discuss
the importance of an initial thermal state for the problem of initial
conditions in ``new inflation'' models with a plateau-like
potential. These  previous works captured much of the important
physics of the problem but were based in several ways on intuitive
reasoning. In the present paper, we will show that
fluctuation-dissipation dynamics during a radiation-dominated
pre-inflationary epoch can indeed lead to such a thermalized state,
where the homogeneous inflaton value naturally becomes localized
within a flat plateau about the origin, thus setting the necessary
conditions for the onset of inflation (see e.g.~\cite{otherpapers} for previous discussions of the role of the pre-inflationary dynamics in setting initial conditions for inflation). 

While the framework of fluctuation-dissipation dynamics allows us to
derive the key results in Ref.~\cite{Albrecht:1984qt}, we will go
further and discuss in detail the conditions under which such a
thermalized state can consistently be reached without spoiling the
inflationary dynamics nor its observational predictions. This, of
course, requires significant interactions between the inflaton and
other degrees of freedom in the pre-inflationary thermal bath, which
is widely believed to be incompatible with successful inflationary
models (see, e.g., Ref.~\cite{Collins:1991gh} for an earlier  account
on this). We wish, in particular, to challenge this prevailing view,
and discuss concrete scenarios where a successful field localization
on the plateau may be achieved.

In addition, we will show that thermal effects may keep playing an
important role even after the inflaton's potential energy becomes
dominant, since for a flat plateau about the origin the field's mass
is dominated by thermal corrections until the temperature drops below
the mass of the thermalized particles with which it interacts. This
leads to a thermal inflation period that follows the early radiation
era but precedes the final slow-roll regime. By lowering the
temperature at which the field starts to move away from the origin,
this period also suppresses nonlinear interactions between the
homogeneous field component and its thermalized sub-Hubble modes, such
that field inhomogeneities play a much less important role than
previously estimated.

{}Finally, we will discuss the possibility of two periods of inflation
occurring in the early Universe. A first period of chaotic inflation
along a monomial branch of the potential can be triggered already at
the Planck scale as discussed above. The energy stored in the inflaton
field at the end of this period still exceeds largely the height of
the plateau, so it is possible that after reheating the inflaton
becomes localized about the origin due to fluctuation-dissipation
effects and a second period of slow-roll eventually begins. The final
50-60 e-folds of inflation will then occur along the plateau-like
region, thus combining the attractive features of both chaotic and new
inflation within the same framework\footnote{For similar ideas using
  an early inflationary epoch that can precede a second inflation
  along a plateau, which could ease the initial conditions problem on
  the latter, see, e.g.,
  Refs.~\cite{Carrasco:2015rva,Artymowski:2016ikw,Dimopoulos:2016yep}. However
  note that among other differences, none of these
 works considered the importance of an intermediary
  reheating phase.}.

This paper is organized as follows. In Sec.~\ref{langevindynamics} we
give the Langevin dynamics describing the evolution of the inflaton
coupled to fermionic fields in the pre-inflationary
thermal bath, when the latter dominate the energy balance. We
explicitly verify that super-Hubble field modes become localized at
the origin if their  decay rate comes to exceed the expansion rate. In
Sec.~\ref{consistency} we analyze the conditions under which this
localization can be achieved, discussing in particular the
compatibility of significant interactions between the inflaton and
other degrees of freedom and quantum corrections in the effective
potential, as well as the role of sub-Hubble modes and the initial
conditions following the Planck era.  In Sec.~\ref{example}, we apply
our results to a concrete example of a plateau-like scalar potential
of the symmetry breaking form, and show explicitly that a period of
thermal inflation may follow the early radiation era. We also discuss the possibility of an early period of chaotic
inflation and the subsequent role of fluctuation-dissipation
dynamics. Our main conclusions are summarized in
Sec.~\ref{conclusions} and we include two appendices with
additional technical details on the dynamics and predictions of
inflation for the concrete example considered and for a
Higgs-like potential.

\section{Dynamical evolution of the inflaton field coupled to a radiation bath}
\label{langevindynamics}

We start by describing the evolution of the inflaton field in a
pre-inflationary era dominated by a radiation bath with which the
inflaton interacts.   We will describe the radiation bath in
terms of a set of $N_F$ fermion species coupled to  the inflaton
through a standard Yukawa coupling of the form:
\begin{equation} \label{yukawa}
\mathcal{L}_Y=-g \phi \sum_{i=1}^{N_F}\bar\psi_i\psi_i.
\end{equation}
The fermions acquire mass through the coupling to the inflaton and
also possibly through other  interactions within the thermal bath.
{}For example, when interacting with abelian gauge bosons with
strength $e$, which keep the fermions in thermal
equilibrium~\cite{bellac} at a temperature $T$,  the fermions can
acquire a thermal plasma mass $m_\psi^2 = e^2 T^2/8$. We will assume
that these masses are negligible with respect to the temperature,
$m_{\psi}\ll T$, but still larger than the Hubble parameter, $m_{\psi}
> H$, such that a flat spacetime quantum field theory calculation can
be used when evaluating the decay width of the inflaton into
fermions. We will also neglect the tree-level inflaton mass, assuming
that the field values of interest correspond to a region of the scalar
potential with negligible curvature compared to the ambient
temperature.   These approximations will need to be checked {\it a
  posteriori} for consistency of the calculation. This is done when we
present an explicit example in Sec.~\ref{example}.

The inflaton nevertheless acquires a thermal mass through the coupling
to the fermions in the radiation bath, which to leading order is given
by~\cite{bellac}
\begin{equation} \label{inflaton_mass}
m_\phi^2=\alpha^2 T^2, \qquad \alpha^2= {g^2N_F\over 6}.
\end{equation}
Interactions with the thermal bath also leads to
fluctuation-dissipation effects. For $m_\phi \gtrsim H$, all
inflaton Fourier modes are oscillating, as we will see below,
and the relevant dissipation coefficient corresponds to the on-shell
decay width into fermions~\cite{Yokoyama:2004pf}, while decay into
bosons (either scalar or gauge) particles can be neglected\footnote{Note that e.g.~for an interaction with scalar bosons of the form $g^2\phi^2\chi^2$, the trilinear vertex leading to the $\phi\rightarrow \chi\chi$ decay is proportional to the tree-level scalar mass $g|\phi| < T$ for relativistic bosons, and this results in a suppression of the bosonic decay channels relative to the fermionic ones by a factor $(g\phi / \alpha T)^2\ll 1$. In any case, the inclusion of additional decay channels will not change our conclusions significantly and in fact it will only enhance the localization effect described below. Additional non-perturbative effects that we are not taking into account in this analysis may also enhance this effect. The same is true if scattering processes are also included. For instance, the scattering processes $\psi_i\bar\psi_i \leftrightarrow \phi\phi$ and $\psi_i \phi \leftrightarrow \psi_i \phi$ correspond to higher-loop contributions to the imaginary part of the inflaton's self-energy. They are thus suppressed compared to the decays considered in this work, which correspond to one-loop processes, and their contribution to the damping rate is further suppressed in the large $N_F$ limit for finite $\alpha$.}. At finite
temperature, for modes of physical three-momentum $p$, the inflaton decay width is given
by~\cite{BasteroGil:2010pb}
\begin{equation} \label{decay_width}
\Gamma_{\phi}(p)={3 \alpha^2 m_\phi^2\over 4\pi
  \omega_p}\left[1+{2T\over p}\ln\left({1+e^{-\omega_+/T}\over
    1+e^{-\omega_-/T}}\right)\right],
\end{equation}
where $\omega_p=\sqrt{p^2+m_\phi^2}$, $\omega_\pm=|\omega_p\pm p|/2$
and $p=|{\bf p}|$ is the modulus of the physical momentum. Since we
will be mainly interested in low-momentum modes, $p\ll T$, and given
that $m_\phi \lesssim T$ for perturbative couplings, we can
approximate the decay width by
\begin{eqnarray} \label{decay_width_approx}
\Gamma_\phi(p) &\simeq & \frac{3 \alpha^2 m_\phi^2}{4 \pi \omega_p}
\left[ 1- 2 \, n_F\left(\frac{m_\phi}{2}\right) \right]
\simeq  \frac{3 \alpha^2 m_\phi^3}{16 \pi \omega_p T},
\end{eqnarray}  
where $n_F$ is the Fermi-Dirac distribution function.
We point out that the additional factor $m_\phi/T< 1$ in the finite temperature decay width with respect to the zero-temperature result is due to Pauli blocking
and reflects the phase space decrease at high temperatures, due to the Fermi-Dirac distribution
in eq.~(\ref{decay_width_approx}). However, since the inflaton mass $m_\phi$ increases with temperature, the decay width is, in fact, larger than at $T=0$. Note that, in a radiation-dominated universe, we have
\begin{equation} \label{radiation_dominated}
a(t)=(t/t_i)^{1/2}, \qquad H={1\over 2t}, \qquad
T(t)=T_i(t/t_i)^{-1/2},
\end{equation}
where we set $a(t_i)=1$ and the initial time is determined by the
Hubble parameter when the initial conditions for the total energy
density are set. This implies that $m_\phi \propto t^{-1/2}$ and since
$p=k/a$ for comoving momentum $k$, we also have $\omega_k\equiv
\omega_p\propto t^{-1/2}$. This yields $\Gamma_\phi \propto
1/\sqrt{t}$, which means that the conditions $\Gamma_\phi,
m_\phi\gtrsim H$ will necessarily be satisfied given a sufficiently
long time. {}For a radiation-dominated universe, the {}Friedmann
relation becomes
\begin{equation} \label{radiation_dominated_hubble}
H=\sqrt{\pi^2\over 90}g_*^{1/2} {T^2\over M_P},
\end{equation}
where $g_*$ denotes the number of relativistic degrees of freedom in
the thermal bath, which includes the fermions and in general other
species not directly coupled to the inflaton and $M_P$ is the reduced
Planck mass, $M_P=2.435 \times 10^{18}$ GeV. Thus, we have in the
limit of vanishing momentum that
\begin{equation} \label{decay_hubble}
{\Gamma_\phi\over H}\simeq 0.18 {\alpha^4\over g_*^{1/2}}{M_P \over
  T},
\end{equation}
which implies that dissipative effects will become significant for
temperatures parametrically below the Planck scale. Since $\Gamma_\phi
\lesssim m_\phi$ for perturbative couplings, the inflaton field will
begin oscillating before dissipative effects become significant.

Assuming a flat FRW geometry, each {}Fourier mode of the inflaton
field will satisfy the Langevin-like equation of the
form~\cite{Berera:1999ws,Berera:2008ar}
\begin{equation} \label{langevin}
\ddot{\phi}_k+(3H+\Gamma_\phi(k))\dot\phi_k+\omega_k^2\phi_k=\xi_k,
\end{equation}
where $\xi_k$ is a stochastic noise term that encodes the backreaction
of the thermal bath on the evolution of the field and is related to
the dissipation coefficient through the fluctuation-dissipation
relation~\cite{Berera:1999ws,Berera:2008ar}
\begin{equation} \label{FD}
\langle\xi_k(t_1)\xi_{k'}(t_2)\rangle=2\Gamma_\phi(k) T {(2\pi)^3\over
  a^3}\delta^3(\mathbf{k+k'})\delta(t_1-t_2).
\end{equation}
This relation is valid in the high-temperature regime, when the fields
coupled to the inflaton are relativistic, and the noise term is approximately
white and Gaussian (with zero average).

To determine the evolution of the Fourier modes, we assume a narrow
decay width $\Gamma_\phi\ll\omega_k$ and large frequency $\omega_k\gg
H$, which should both be valid when the couplings are small and $T\gg
H$. The equation can be put into a  suitable form for a JWKB analysis
by introducing $\tilde\phi_k$ from
\begin{equation}
\phi_k=a^{-3/2}e^{-\Gamma_\phi/(2H)}\tilde\phi_k \;.\label{fouriermode}
\end{equation}
The approximations mentioned above now imply that
\begin{equation}
\ddot{\tilde\phi}_k+\omega_k^2\tilde\phi_k\approx
\tilde\xi_k=a^{3/2}e^{\Gamma_\phi/(2H)}\xi_k \;.
\end{equation}
The JWKB solutions to the homogeneous equation are
\begin{equation}
\tilde\phi_k^\pm\approx \omega_k^{-1/2}e^{\pm i\int\omega_k dt}.
\end{equation}
The solutions (\ref{fouriermode}) then yield oscillating field modes,
the amplitude of which redshifts with expansion and also decays
exponentially in time due to dissipative effects. As seen above, the
latter will become significant at parametrically low temperatures and
we would expect the field to completely decay away given enough
time. This is analogous to the damping of a Brownian particle's motion
in a molecular gas due to the average friction effect resulting from
random collisions with the molecules in the
environment~\cite{brownian}.  

The JWKB solutions provide the Green functions which we can use to
solve the  inhomogeneous equation,
\begin{equation}
\tilde\phi_k(t)=\int_{t_i}^t\,\tilde
G(t,s)\tilde\xi_k(s),\label{particular_sol}
\end{equation}
where
\begin{equation}
\tilde G(t,s)\approx
\omega_k(t)^{-1/2}\omega_k(s)^{-1/2}\sin\int_s^t\omega_k(t')dt'.
\end{equation}
The quantity of interest is the power spectrum of fluctuations,
\begin{equation}
\langle\tilde\phi_k(t)\tilde\phi_{k'}(t)\rangle=\int_{t_i}^t
ds_1\int_{t_i}^t ds_2\, \tilde G(t,s_1)\tilde
G(t,s_2)\langle\tilde\xi_k(s_1)\tilde\xi_k(s_2)\rangle.
\end{equation}
Using the noise correlation relation Eq.~(\ref{FD}), and restoring the
original field modes, we may write the power spectrum as
\begin{equation}
\langle\phi_k(t)\phi_{k'}(t)\rangle=2e^{-\Gamma_\phi/H(t)}a(t)^{-3}\int_{t_i}^t
           {ds\over \omega_k(s)\omega_k(t)}\,
           \left[\sin^2\int_s^t\omega_k(t')dt'\right]\Gamma_\phi(s)T(s)e^{\Gamma_\phi/H(s)}.
\end{equation}
Now a remarkable simplification occurs for the dissipation term given
above. We notice that  $\Gamma_\phi /H\propto a$, and consequently
\begin{equation}
{\Gamma_\phi T\over \omega_k}ds \propto da.
\end{equation}
The equal-time correlator can then be obtained by averaging over the
field oscillations (using that $\langle\sin^2x\rangle=1/2$) and
integrating a simple exponential to give
\begin{eqnarray} \label{correlator_k_t}
\langle \phi_k (t)\phi_{k'}(t)\rangle &\simeq &{T(t)\over
  \omega_k^2(t)
  a^{3}(t)}\left[1-e^{-[\Gamma_\phi/H(t)-\Gamma_\phi/H(t_i)]}\right](2\pi)^3
\delta^3(\mathbf{k}+\mathbf{k'})
\nonumber\\  &\simeq&
  {T(t)\over
  \omega_k^2(t) a^{3}(t)}(2\pi)^3\delta^3(\mathbf{k}+\mathbf{k'}),
\end{eqnarray}
where we have taken $\Gamma_\phi \gtrsim H$ in the second equality. 

It is important to point out that the result (\ref{correlator_k_t})
applies not only to sub- but also to super-Hubbles modes in the
present set up. A priori one 
should expect super-Hubble modes not to be thermal. They should be 
frozen and maybe squeezed, as super-Hubble modes of cosmological 
perturbations are~\cite{Mukhanov:1990me}. However, we are considering the radiation
epoch, and in this epoch super-Hubble fluctuations are not frozen out and are
not squeezed. The reason is that the usual
squeezing factor $a''/a$ vanishes. In
addition, we are considering a massive
field, and so the super-Hubble modes will
be oscillating and their evolution will
remain adiabatic. In this way, the fluctuation
term conveys thermal equilibrium amplitude
values to super-Hubble modes.

Note that the power spectrum redshifts as $t^{-1}$ in the radiation
dominated era, which is the regime we assume to be most relevant for
the pre-inflationary era.  The result still applies, however, when
radiation and potential energy terms are comparable, and we shall need
this later.  Note also that the field variance becomes independent of
the friction coefficient  for $\Gamma_\phi \gtrsim H$. Thus, the noise
term counteracts the effects of dissipation,  making the field modes
evolve towards an equilibrium configuration determined by the ambient
temperature $T(t)$. This would follow from local flatness and the
equivalence principle for scales smaller than the  horizon (see, e.g.,
Ref.~\cite{Bastero-Gil:2014jsa}), but the important  point here is
that the result holds on long as well as short scales. Again, this  is
analogous to the motion of a Brownian particle described by the
Langevin equation, where the random effect of molecular collisions
does not really lead to the complete damping of the particle's motion
through dissipative friction but rather to thermal fluctuations of the
particle's velocity in an equilibrium configuration~\cite{brownian}.

We may now use the result given by Eq.~(\ref{correlator_k_t}) to
determine the average field value in configuration space. We are
mainly interested in the ``classical" or ``homogeneous" field
component, since only the low-momentum or superhorizon modes, $k<aH$,
can lead to a slow-roll inflationary regime, while sub-horizon modes
will redshift away more quickly~\cite{Kung:1989xz,
  Brandenberger:1990wu, Brandenberger:1990xu}. Since the field
fluctuations in a thermal bath are classical in nature, we may use the
{}Fourier mode decomposition for a real classical field,
\begin{eqnarray} \label{Fourier}
\phi(\mathbf{x},t)=\int {d^3k\over
  (2\pi)^3}\phi_k(t)\cos\left(\mathbf{k}\cdot\mathbf{x}+\alpha_{\mathbf k}\right)~,
\end{eqnarray}
where $\alpha_{\mathbf{k}}$ are mode-dependent phases. Assuming the latter to be randomly distributed and thus averaging to zero, we may then write the average field value on superhorizon scales as:
\begin{eqnarray} \label{classical_field}
\langle \phi_c^2(\mathbf{x}, t)\rangle &=& \int_{k,k'<aH}{d^3k\over
  (2\pi)^3}{d^3k'\over (2\pi)^3}\langle \phi_k(t)\phi_{k'}(t)\rangle
\cos\left(\mathbf{k}\cdot\mathbf{x}\right)\cos\left(\mathbf{k'}\cdot
\mathbf{x}\right)\nonumber\\ &\simeq & {T\over 4\pi^2
  a^{3}}\int_{0}^{aH}{k^2dk\over \omega_k^2}\int_{-1}^{1}d\cos\theta
\cos^2(kr\cos\theta)\nonumber\\ &\simeq & {T\over 4\pi^2
  a^3}\int_{0}^{aH}{k^2dk\over \omega_k^2}\left[1+{\sin(2kr)\over
    2kr}\right].
\end{eqnarray}
The oscillatory term between brackets in the last line in
Eq.~(\ref{classical_field}) makes the exact integration rather
cumbersome, but since it is roughly between 1-2, we may obtain a good
estimate for the field variance by replacing it with an
$\mathcal{O}(1)$ number, say 2 for concreteness and to give the
largest possible variance. This then yields the result
\begin{eqnarray} \label{classical_field_bound}
\langle \phi_c^2\rangle &\sim  & {T\over 2\pi^2
  a^3}\int_{0}^{aH}{k^2dk\over \omega_k^2} \sim 
        {H^3T\over 6\pi^2 m_\phi^2} \sim  {1\over 6\pi^2
          \alpha^2}{H^3\over T},
\end{eqnarray}
where we have used that $m_\phi\gtrsim H$ in the pre-inflationary era. This implies that the
mean-square classical field fluctuation decreases as $t^{-5/2}$ and we
may write it only in terms of the temperature of the thermal bath to
finally obtain the result
\begin{eqnarray} \label{classical_field_T}
\langle \phi_c^2\rangle & \sim & {\pi\over 6\alpha^2}\left({g_*\over
  90}\right)^{3/2}{T^5\over M_P^3}.
\end{eqnarray}
{}From the results given by Eqs.~(\ref{classical_field_bound}) and
(\ref{classical_field_T}),  we thus conclude that, asymptotically, the
average inflaton field variation is below the ambient temperature and
also below the Hubble scale since $H^3/T  < H^2$. This holds when the
effective coupling is not too small, but notice that we required
$m_\phi\simeq \alpha T \gtrsim H$ for consistency of the calculation.

\section{General Consistency Conditions Analysis}
\label{consistency}

The analysis given in the previous section shows that
fluctuation-dissipation  effects can localize the homogeneous inflaton
field at the origin with a parametrically small dispersion
$\Delta\phi_c/H \sim 0.1\alpha^{-1} (H/T)^{1/2}$, as inferred from the
result in Eq.~(\ref{classical_field_bound}). This will occur in a
pre-inflationary radiation-dominated epoch provided that the inflaton
field interacts significantly with the ambient radiation, modeled by a
set of fermion species in our analysis. We are interested in scalar
potentials that have a flat plateau about the origin, where the
curvature is negligible and the inflaton mass is dominated by thermal
effects. We will consider a specific example of such a potential in
Sec.~\ref{example}, but in general we may consider a flat plateau
region of height $V_0$. Thus, when the field is localized in this
plateau, the inflaton potential will dominate over the energy density
in radiation at temperatures below the {\it equality temperature}
$T_{eq}$. This can be obtained by equating $\rho_R(T_{eq})= V_0$,
which yields:
\begin{eqnarray} \label{T_critical}
T_{eq} = \left({30\over \pi^2}\right)^{1/4} g_*^{-1/4}V_0^{1/4}.
\end{eqnarray}
We may relate the height of the plateau with the tensor-to-scalar
ratio and the known amplitude of scalar curvature perturbations,
$\Delta_\mathcal{R}^2\simeq 2.2\times 10^{-9}$ \cite{Ade:2015lrj},
generated in the slow-roll regime,
\begin{eqnarray} \label{rT_critical}
r={2\over 3\pi^2}\Delta_\mathcal{R}^{-2}{V_0\over M_P^4}.
\end{eqnarray}
Using Eq.~(\ref{decay_hubble}), one may then obtain a lower bound on
the effective coupling $\alpha$, defined in Eq.~(\ref{inflaton_mass}),
which is required such that $\Gamma_\phi>H$ before the transition
temperature for vacuum dominance over radiation is reached and the
pre-inflationary radiation era comes to an end. This yields
\begin{eqnarray} \label{bound_coupling}
\alpha\gtrsim 0.24 g_*^{1/16}\left({r\over 10^{-6}}\right)^{1/16}.
\end{eqnarray}
Hence, the coupling between the inflaton and the thermal bath has to
be quite large unless inflation occurs at very low scales and yields a
negligible tensor-to-scalar ratio. Otherwise, dissipative friction
will not have enough time to damp any initially large field values and
stabilize the variance at the value computed in the previous section.
Such coupling values may be too large to yield a successful
inflationary  model, since radiative corrections to the scalar
potential may spoil its flatness and fail to give the observed
amplitude of the primordial curvature perturbation spectrum. This
requires further examination of the effective potential during
inflation. Likewise, we have to make sure that the energy density in
the inhomogeneous (sub-Hubble) inflaton modes remains sub-dominant, so
as not to spoil the onset of accelerated expansion. Finally, we also
need to ensure that, as assumed, the fermions remain light throughout
the whole of the pre-inflationary era, which is crucial for the
thermalization and consequent localization of the inflaton
field. Below we analyze each of these issues in more detail.

\subsection{Radiative Corrections and the Flatness of the Inflaton Potential}
\label{decoupling}

As shown above, the coupling of the inflaton to the radiation bath
fields, Eq.~(\ref{bound_coupling}), has to be rather large so as to
allow a sufficiently fast localization of the inflaton field with
small variance. This in turn can potentially lead to large radiative
corrections to the inflaton potential, possibly endangering its
flatness. Two possible solutions to this problem, and that we discuss
next, can be given in terms of inflation models based on
supersymmetry, or, alternatively, by the appropriate application of
the renormalization procedure determining these radiative corrections.
 
In supersymmetric models of inflation, large (although perturbative)
couplings may be allowed.  Despite supersymmetry being broken by the
finite energy density during inflation, there is still a  partial
cancellation of bosonic and fermionic radiative corrections. {}For
example, for a superpotential  of the form $W=g\Phi X_i^2/2+f(\Phi)$,
which gives a Yukawa term of the form in Eq.~(\ref{yukawa}),  if the
inflaton is the scalar component of the $\Phi$ supermultiplet and the
fermions are part of the  chiral multiplets $X_i$, $i=1,\dots, N_F$,
one obtains the one-loop correction to the scalar
potential~\cite{BasteroGil:2012cm}:
\begin{eqnarray} \label{CW_SUSY}
\Delta V^{(1)}_{SUSY}(\phi)={3\alpha^2\over
  8\pi^2}\ln\left({g^2\phi^2\over\mu^2}\right)V(\phi),
\end{eqnarray}
where $\mu$ is the renormalization scale in the
$\overline{\mathrm{MS}}$ scheme, and $V(\phi)=|f'(\phi)|^2$ is the
tree-level scalar potential,  such that one-loop corrections give
naturally a sub-dominant contribution to the effective potential and
do not significantly modify the tree-level predictions. Note that we
may easily construct a supersymmetric version of the analysis
performed in the previous section, with bosonic fields contributing to
the inflaton's thermal mass and decay width. The contribution of
bosons to the latter is, in fact, suppressed at high temperatures as
we have already discussed, while the thermal mass contribution is
analogous to that of the fermions up to numerical
factors. Supersymmetry can then allow for the significant coupling
between the inflaton and a radiation bath required for the
localization of its homogeneous modes close to the origin without
considerably changing the form of the potential at zero-temperature. 

Alternatively, it has also been argued that, if the particles coupled
to the inflaton are heavy compared to  the relevant energy scale $\mu$
during inflation, their contribution to the effective potential is
further suppressed,  in the spirit of the decoupling
theorem~\cite{Appelquist:1974tg}.  This becomes explicit using a
mass-dependent renormalization scheme~\cite{BasteroGil:2010vq},
whereas in  mass-independent schemes, such as the modified minimal
subtraction $\overline{\mathrm{MS}}$ scheme used in dimensional
regularization, it has to be enforced by hand. In particular, in a
mass-dependent renormalization scheme,  the one-loop contribution of
fields coupled to the inflaton is given, at one-loop order, by
\begin{eqnarray} \label{CW_MD}
\Delta V^{(1)}_{MD}(\phi)= {g^4N_F\phi^4\over
  64\pi^2}\left[\ln\left({M^2\over \mu^2}\right)- I\left({M^2\over
    \mu^2}\right)\right],
\end{eqnarray}
where the function
$I(a)=\ln(a)-2-\sqrt{1+4a}\ln\left[(\sqrt{1+4a}-1)/(\sqrt{1+4a}+1)\right]$
effectively replaces the $3/2$ factor that one obtains in the
$\overline{\mathrm{MS}}$ regularization scheme. This has been derived
in Ref.~\cite{BasteroGil:2010vq}  for a scalar interaction
$g^2\phi^2\chi^2/2$, yielding an effective mass squared for the $\chi$
particles given by $M^2=m_\chi^2+g^2\phi^2$. An analogous  result can
also be derived for fermions, which gives an expression of the form of
Eq.~(\ref{CW_MD}), but with an opposite sign. {}For heavy fields,
with $\mu \ll M$, Eq.~(\ref{CW_MD}) becomes
\begin{eqnarray} \label{CW_MD_approx}
\Delta V^{(1)}_{MD}(\phi)\simeq - {g^4N_F\phi^4\over
  384\pi^2}{\mu^2\over M^2}.
\end{eqnarray}
In particular, if the field-independent mass in
$M^2=m_\chi^2+g^2\phi^2$ can be neglected, we immediately see that
this  correction to the potential is quadratic in $\phi$, thus
shifting its squared mass by an amount
$(\alpha^2/32\pi^2)\mu^2$. Arguably the relevant energy scale during
inflation is the Hubble scale,  $\mu\sim H_{inf}$, such that this correction
to the inflaton mass is small even for the relatively large values of
the effective coupling $\alpha$ required for thermalization, and does
not spoil the flatness of the potential. Note that the temperature during the pre-inflationary radiation era $T>H>H_{inf}$, so that we may have the hierarchy $T> M>H_{inf}$ for which fields in the thermal bath are relativistic before inflation but decouple during the slow-roll phase.  The appropriate choice for
the renormalization scale during inflation remains, however, the
subject  of some discussion in the literature, but it is nevertheless
clear that there are possible scenarios  where the inflaton can couple
sufficiently strongly to a pre-inflationary thermal bath  without
spoiling the required properties of the scalar potential at the quantum
level.


\subsection{Kinetic and gradient energies before inflation}

We have concluded in the previous section that the super-Hubble (or
homogeneous) inflaton modes become localized at the origin due to
fluctuation-dissipation effects. We need, however, to make sure that
(i) their velocity and spatial variation are also kept small and (ii)
the energy density stored in higher-momentum modes is diluted fast
enough compared to the potential energy. These conditions are crucial
for the homogeneous field to enter a slow-roll regime.

As we have seen earlier in the previous
Section~\ref{langevindynamics}, provided that $\Gamma_\phi >H$, the
inflaton field coupled to the radiation bath will tend to an
equilibrium solution characterized by a ``classical" component of
average  $\langle \phi_c^2\rangle\sim H^3T /6\pi^2 m_\phi^2$, with
$m_\phi^2\simeq \alpha^2T^2$.  {}From the inhomogeneous solution we
have obtained in that Section, we may also compute the  average field
velocity and gradients, in order to better understand whether the
field localization  achieved by interactions with the thermal bath
leads to appropriate initial conditions for inflation  with a
plateau-like potential.

{}First, differentiating Eq.~(\ref{particular_sol}) with respect to
time, again using the JWKB approximation, we obtain:
\begin{equation} \label{particular_sol_dot}
\dot{\tilde\phi}_k\approx\omega_k(t)\int_{t_i}^t\left[\int_s^t\cos\omega_k(t')dt'\right]
\tilde\xi_k(s){ds\over\sqrt{\omega_k(s)\omega_k(t)}}.
\end{equation}
We may then proceed as for the computation of the field variance to
obtain the variance of the field velocity.  In this case the main
differences are the factors outside the integral and the fact that the
integral  includes a cosine function rather than a sine. Since we are
averaging over field oscillations,  the latter difference is
irrelevant and it is straightforward to obtain after some algebra the
result
\begin{eqnarray} \label{velocity_correlator_k_t}
\langle \dot{\phi}_k (t)\dot{\phi}_{k'}(t)\rangle &\simeq & {T(t)\over
  a^{3}(t)}(2\pi)^3\delta^3(\mathbf{k}+\mathbf{k'}) =
\omega_k^2(t) \langle \phi_k(t)\phi_{k'}(t)\rangle.
\end{eqnarray}
The above result is what one would expect for an oscillating field
with frequency $\omega_k$.  {}For small momentum modes such that
$p=k/a \ll m_\phi \sim \alpha T $, this frequency is simply the
field's thermal mass,  such that it is easy to obtain the average
squared field velocity for the classical component,  including all
comoving momenta up to $k=aH$,
\begin{eqnarray} \label{average_velocity_classical}
\langle \dot{\phi}_c^2\rangle \simeq m_\phi^2 \langle \phi_c^2\rangle
\simeq {H^3 T\over 6\pi^2}.
\end{eqnarray}
This implies that the average kinetic energy in the classical field
decays as $a^{-7}$ during the pre-inflationary  radiation era, and so
much faster than the radiation fluid itself. 

The above calculation also provides us with the initial conditions in
phase space for the classical  (homogeneous) field at the equality
temperature, given in Eq.~(\ref{T_critical}). This yields
\begin{eqnarray} \label{initial_cond_phase_space1}
\phi_c &\sim & \sqrt{\langle \phi_c^2\rangle} 
\simeq 0.08 {g_*^{1/8}\over \alpha}  \left({V_0\over M_P^4}\right)^{5/8} M_P
 \simeq  10^{-7}  {g_*^{1/8}\over \alpha}\left({r\over
      0.01}\right)^{5/8}M_P,
\end{eqnarray}
\begin{eqnarray} \label{initial_cond_phase_space2}
 \phi'_c & \sim  & H \sqrt{\langle
      \dot{\phi}_c^2\rangle}   
 \simeq  0.13 g_*^{-1/8}
    \left({V_0\over M_P^4}\right)^{3/8}  M_P
 \simeq  4\times 10^{-5}
    g_*^{-1/8}\left({r\over 0.01}\right)^{3/8}M_P.
\end{eqnarray}
Notice that this is still far from a slow-roll solution, since
$\phi'_c/\phi_c \sim \alpha T/H \gg 1$,  which is naturally expected
since the field is oscillating. We will nevertheless show in the next
section, where we consider a concrete example of a plateau-like
potential, that these initial conditions can yield a sufficiently long
period of inflation in the slow-roll regime. Note also that the recent
Planck results~\cite{Ade:2015lrj} yield $H_{\rm inf} \lesssim 10^{13}$
GeV for the Hubble parameter during inflation, such that $V_0\ll
M_P^4$.  Thus, both the average field value and the average field
velocity are (very) sub-Planckian.  Note, in addition, that the
average super-horizon field variance can be written as
\begin{eqnarray} \label{variance_quantum}
\langle \phi_c^2\rangle \simeq {2\over 3\alpha^2} \left({H\over
  T}\right)  \left({H\over 2\pi}\right)^2.
\end{eqnarray}
Thus, in the presence of the thermal bath the average field
fluctuations on super-horizon scales are smaller  than the quantum
fluctuations in the de Sitter inflationary phase on the scale of the
horizon with amplitude $\sim H/2\pi$.

We may proceed analogously to compute the gradient energy, noting that
\begin{eqnarray} \label{gradient_squared}
\langle \left(\nabla \phi(\mathbf{x},t)\right)^2\rangle &=& \int
        {d^3k\over (2\pi)^3} \int {d^3k'\over
          (2\pi)^3}\mathbf{k}\cdot\mathbf{k'}\langle
        \phi_k(t)\phi_{k'}(t) \rangle \sin
        (\mathbf{k}\cdot\mathbf{x})\sin (\mathbf{k'}\cdot\mathbf{x})
        \nonumber\\ &=& {T\over a^3}\int {d^3k\over (2\pi)^3}
                    {k^2\over \omega_k^2}
                    \sin^2(\mathbf{k}\cdot\mathbf{x}) \nonumber\\ &=&
                        {T\over 4\pi^2a^3}\int {k^4 dk\over
                          \omega_k^2} \left[1-{\sin(2kr)\over
                            2kr}\right],
\end{eqnarray}
where in the last line we have performed the integral over the angular
variables. The term within square brackets is $\mathcal{O}(1)$ and we
may  replace it by unity to estimate the gradient energy.  {}For the
classical field component, we include only the  super-horizon modes in
the momentum integral, which then gives  for the gradient energy the
result
\begin{eqnarray} \label{gradient_squared_2}
{1\over 2a^2}\langle (\nabla \phi_c)^2\rangle \simeq {1\over 40\pi^2
  \alpha^2} {H^5\over T},
\end{eqnarray}
 which decays as $a^{-11}$. Note that this is essentially $ H^2\langle
 \phi_c^2\rangle $ up  to numerical factors. Hence,  $\langle \phi_c^2
 \rangle$ gives the mean variation of the field over super-horizon
 distances.
 
In the calculations shown above we have considered only the classical
field component, including only  super-horizon momentum modes, since
these are the only ones that may eventually enter a slow-roll  regime
after the radiation-dominated period has ended. It is nevertheless
relevant to compute the  average values of the different energy
components when including higher-momentum modes.  Our analysis is
valid up to physical momenta of the order of the ambient temperature,
$p=k/a \lesssim T$,  for which $\omega_k \sim \alpha T$, for
$\alpha\lesssim 1$. Higher-momentum modes will naturally  be
(Boltzmann) suppressed in the thermal bath, such that we may safely
neglect them. Thus, extending our earlier  calculations to $k<aT$, we
obtain that the average potential $\langle V\rangle$, field velocity energy density
$\langle \rho_{\dot\phi}\rangle$ and the energy density in the field
gradient $\langle \rho_\nabla\rangle$ are given, respectively, by 
\begin{eqnarray} \label{full_energies1}
\langle V\rangle & = &{1\over 2} m_\phi^2 \langle\phi^2\rangle \simeq
        {1\over 4\pi^2} T^4, \\  
\nonumber \\
\langle
        \rho_{\dot\phi}\rangle&=&{1\over 2} \langle
        \dot{\phi}^2\rangle \simeq {1\over 4\pi^2}  T^4,
        \label{full_energies2} \\  
\nonumber \\
\langle \rho_\nabla\rangle &=& {1\over
          2a^2}\langle  (\nabla \phi)^2\rangle \simeq {1\over 40\pi^2
          \alpha^2} T^4.
\label{full_energies3}
\end{eqnarray}
This shows that the potential, the kinetic and the gradient energies
in sub-horizon modes are diluted like radiation.  This should be
expected, since for $\Gamma_\phi >H$ the inflaton field thermalizes,
and its short-wavelength  components should behave as relativistic
degrees of freedom in the thermal bath.  The numerical factors in
Eqs.~(\ref{full_energies1}), (\ref{full_energies2}) and (\ref{full_energies3})  
are not rigorous, due to the approximations
considered when estimating them, but this shows that essentially the
inflaton field can be split into two separate components: the
long-wavelength modes,  which may eventually enter the slow-roll
regime; and the short-wavelength modes, which are essentially  part of
the thermal bath and should correspond to an additional relativistic
degree of freedom  that must be taken into account when computing
$g_*$. Once the radiation energy density $\rho_R$ satisfies $\rho_R<
V_0$, the long-wavelength modes will dominate the evolution if they
enter a slow-roll  regime, according to the discussion above, and
neither the kinetic nor the gradient energies in  the short-wavelength
components can block inflation, since, like radiation, they will be
diluted away.

Notice that one could worry that, even if the field's gradient energy
is negligible,  large fluctuations in the field could bring the
classical component outside the plateau  where slow-roll is possible
through nonlinear effects. The analysis above shows that  the largest
fluctuations of the field occur on sub-horizon scales of size $\sim
T^{-1}$  and with amplitude $\sim T$. This would imply that the
plateau of the inflaton potential must  in general
extend for  $\Delta \phi \gtrsim T_{eq}\sim V_0^{1/4}$ about the
origin, i.e., that the plateau's width must be at least comparable to
its height. However, the inflaton's self-interactions are typically
extremely weak and, as we will discuss in the next section with a
concrete example, slow-roll does not start immediately at the equality
temperature. We will thus leave the study of nonlinear mixing between
sub- and super-Hubble modes to the next section.

\subsection{Thermalization and equipartition of energy}

Our analysis of the thermalization process prior to the inflationary
regime relies on the assumption that the inflaton is coupled to
thermalized light particles that dominate the energy balance before
the onset of inflation. If this is the case, the inflaton field will
be localized at the origin by thermal and fluctuation-dissipation
effects as we have demonstrated, but we need to check the consistency
of this assumption.

Since our analysis is based on general relativity and perturbative
quantum field theory, it will only hold for energies below the Planck
scale. We must then ask what are the reasonable initial conditions as
the Universe emerges from the Planck era, when quantum gravity effects
are presumably dominant. In principle, the Universe emerges from this
era, i.e., quantum gravity effects become sub-leading, when the total
energy density becomes sub-planckian, i.e., when $\rho_{T}\sim
M_P^4$. Since gravity treats all species on equal footing, it is also
reasonable to assume equipartition of energy after the Planck era. We
assumed that there are $g_*$ relativistic degrees of freedom at this
stage (and throughout the pre-inflationary stage) that quickly
thermalize, and the inflaton field should thus carry as much energy
density as each of these degrees of freedom. Thus, for $g_*\gg 1$, we
have that
\begin{eqnarray} 
\rho_T \sim  g_*T^4 + \rho_\phi \sim M_P^4\qquad \Rightarrow \qquad
\rho_\phi \sim T^4\sim g_*^{-1} M_P^4.
\end{eqnarray}
Hence, we expect that immediately after the Planck era the Universe
will have a temperature parametrically below $M_P$ and the energy in
the inflaton field is also parametrically sub-planckian. Nevertheless,
for a not too large number of relativistic degrees of freedom, we
expect that at this stage $\rho_\phi \gg V_0$ and that the temperature
is well above the equality temperature defined earlier, $T_{eq}\sim
g_*^{-1/4} V_0^{1/4}$.  

Consistency of the analysis requires, in particular, that the fermions
are light, $g|\phi|\lesssim T$. If this is the case, our analysis
shows that the field amplitude redshifts as $1/a$ and is further
reduced by fluctuation-dissipation effects through the inflaton decay
into fermions. Since $T\propto 1/a$, this implies that if the
condition $g|\phi|<T$ is satisfied at the Planck era, it will be
satisfied at all times afterwards. Thus, we need to check under which
conditions one can have light fermions already at the Planck era, and
analyze what happens otherwise.

To properly address the above consistency conditions  for the
localization of the inflaton field, it is more convenient to show how
they work for an explicit example, as we will do in the following section.


\section{Initial Condition Problem for Plateau-Like Inflaton Potentials:
An explicit example}
\label{example}

The main constraints on any inflaton potential are that it must lead
to enough inflation, with number of e-folds of expansion
around or greater than  $N_e \sim 50-60$, to give the correct amplitude of
scalar curvature perturbations, $\Delta_\mathcal{R}^2\simeq 2.2\times
10^{-9}$, and to produce satisfactory values for the observational
quantities, like the tensor-to-scalar ratio $r$ and the spectral index
$n_s$, which according to the recent Planck data~\cite{Ade:2015lrj}
are constrained to the values $r < 0.08$ ($95\%$ CL, TT+lowP+BKP) and
$n_s=0.9655\pm 0.0062$ ($68\%$ CL, Planck TT+lowP), respectively.

The upper bound on the tensor-to-scalar ratio, in particular, implies that
large-field cold inflation models, 
such as monomial potentials, are generically
disfavored by the data, which thus seems to prefer scalar potentials
with a plateau-like region. We are mainly interested in potentials for
which the plateau is centered at $\phi=0$, in which case
fluctuation-dissipation effects may localize the field in the plateau
during a pre-inflationary radiation-dominated era as we have discussed
in the previous sections. This suggests potentials with a
$\mathbb{Z}_2$ reflection symmetry with minima displaced from the
origin at $\phi=\pm v$. The simplest example of such a potential is,
of course, the Higgs type of symmetry breaking potential 
$V(\phi)= \lambda(\phi^2-v^2)^2/4$, but
it is easy to check that such a potential only admits slow-roll
solutions for $v\gtrsim M_P$. These solutions involve super-planckian
field values, and whether this poses a problem for inflationary
model-building has been the object of much discussion in the
literature. Nevertheless, independently of this fact, when slow-roll
solutions exist they are phase-space attractors, as we explicitly show
in Appendix~\ref{Sec:Higgs}. In this sense there is no real initial conditions
problem for this renormalizable double-well potential, and it behaves
very much like a large-field model, although it may lead to a
tensor-to-scalar ratio within observational bounds.

\begin{figure}[htbp]
\centering\includegraphics[scale=1.2]{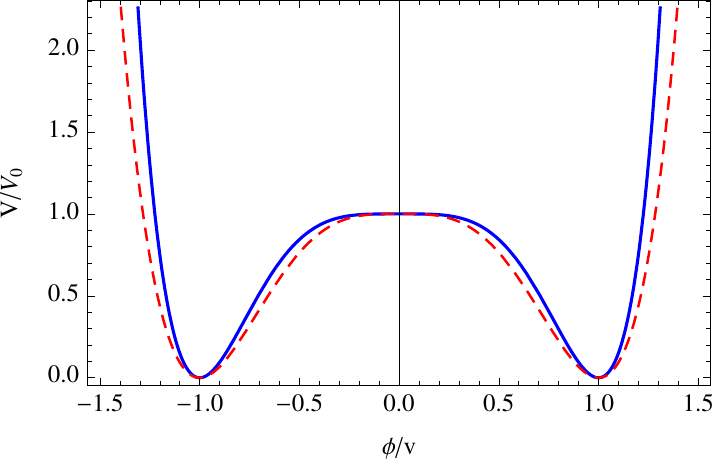}
\caption{Non-renormalizable plateau-like potential (solid blue curve)
  and Coleman-Weinberg potential (dashed red curve).}
\label{sextic:fig}

\end{figure}

A fine-tuning of initial conditions is only required in cases where
slow-roll solutions are not phase-space attractors, and this occurs
for plateau-potentials about the origin such that
$V'(0)=V''(0)=0$. This is the case of the Coleman-Weinberg potential
$V=V_0\left[1+x^4(2\ln x^2-1)\right]$, where $x=\phi/v$, that results from
one-loop quantum corrections to a classically flat potential and, thus,
yields the well-known radiative symmetry breaking mechanism. It is also the case
of symmetry-breaking potentials with non-renormalizable terms, such
as:
\begin{eqnarray} \label{sextic}
V(\phi)= V_0 \left(1-3 \frac{\phi^4}{v^4} +2
\frac{\phi^6}{v^6}\right), 
\end{eqnarray}
where $V_0 = \lambda v^4$. We plot these two potentials in Fig.~\ref{sextic:fig}, where the similarities between them are apparent. In fact, they
lead to quite similar observational predictions for $n_s$ and $r$ as
can be easily checked.

 {}For simplicity of the calculations, and despite its
 non-renormalizable nature, we will take the latter as the particular
 example of a plateau-like potential for which we will explicitly
 check the efficiency and validity of the generic localization
 mechanism through fluctuation-dissipation effects. The subsequent
 discussion will nevertheless apply in a similar way to the
 Coleman-Weinberg case and other similar plateau-like models such as
 the generic non-renormalizable models described in Appendix~\ref{plateauV}. 
 
 We note that with potentials of this form one can obtain
 (non-attractor) slow-roll solutions for sub-planckian field values,
 as explicitly shown in Appendix~\ref{plateauV}. This leads to low values of
 the tensor-to-scalar ratio, possibly outside the reach of near-future
 experiments, and for the non-renormalizable potential in
 Eq.~(\ref{sextic}) one obtains in this case $n_s\simeq
 0.94-0.95$. Although this is outside the window allowed by the latest
 Planck data, we point out that non-renormalizable plateau-like
 potentials with higher field powers yield larger values for $n_s$,
 within the observationally allowed window, and that even for the case
 in study the addition of dissipative effects during inflation
 (leading to a warm inflation regime) can sufficiently increase the
 spectral index (see, e.g., Ref.~\cite{Benetti:2016jhf}). We leave 
some of the details of the dynamics of inflation and
 observational predictions for these potentials to the appendix, and
 here we focus on the problem of initial conditions, which is the main
 scope of this work.

We have studied the dynamics of inflation for the non-renormalizable
plateau potential, in particularly solving numerically the coupled
inflaton and Friedmann equations so as to determine the initial
conditions in phase space leading to a sufficiently long period of
inflation. From this study, detailed in Appendix~\ref{plateauV}, we have obtained
the following constraints on the initial field value and velocity,
respectively,
\begin{eqnarray} \label{NR_phiphil1}
|\phi_i|/M_P &\lesssim & 3\times 10^{-2} (v/M_P)^2, 
\\   
\nonumber \\
|\phi_i'|/M_P &\lesssim&  8\times 10^{-2} (v/M_P)^2,
\label{NR_phiphil2}
\end{eqnarray}
where $\phi'\equiv d\phi/dN_e$. Taking into account that for this
model $r\simeq 4\times 10^{-7}(v/M_P)^4$ for $v\lesssim M_P$, the
average values obtained for the field and velocity at the temperature
$T_{eq}$, where the inflaton potential dominates over the radiation
energy density, and given in Eqs.~(\ref{initial_cond_phase_space1})
and (\ref{initial_cond_phase_space2}), can be
written, respectively, as
\begin{eqnarray} \label{initial_cond_phase_space_NR}
\phi_c \simeq 2\times 10^{-10} {g_*^{1/8}\over \alpha}\left({v\over
  M_P}\right)^{5/2}M_P,  \\
\nonumber\\  \phi'_c \simeq 9\times 10^{-7}
g_*^{-1/8}\left({v\over M_P}\right)^{3/2}M_P.
\end{eqnarray}
This shows that, with the pre-inflationary thermalization period, the
average field value is always within the required range of initial
conditions for any $v<M_P$, while the field velocity is within the
required range for $10^{-8}M_P\lesssim v <M_P$. There is, thus, a wide
parametric range for which inflation starts naturally after the
initial radiation-dominated period. 

We also note that the amplitude of the primordial curvature power
spectrum $\Delta_\mathcal{R}^2\simeq (2/3\pi^2)(V_0/M_P^4)r^{-1}\simeq
2.2\times 10^{-9}$ yields $\lambda = V_0/v^4 \simeq 10^{-14}$, which
is the typical value for inflaton self-interactions found generically
in single-field inflation models. We will use this below to estimate
the effects of nonlinearities in the field potential on the dynamics
of the long-wavelength modes. 

One aspect that we may immediately check is whether our assumption
that the fermions coupled to the inflaton remain light throughout the
initial radiation-dominated era is satisfied. If one assumes
equipartition of energy at the end of the Planck era, as discussed
earlier, we have an initial temperature $T\sim g_*^{-1/4}M_P $
parametrically below the Planck scale, and $V(\phi)\sim T^4 \sim
M_P^4/g_*$. In the presence of a thermal bath of relativistic
particles, the leading thermal correction to the inflaton potential is
the thermal mass term $\alpha^2 T^2 \phi^2$, as discussed earlier, and
one can easily check that for $\lambda\sim 10^{-14}$ this term
dominates over the zero-temperature part of the potential at such a
high temperature for field values such that $V(\phi) \sim
M_P^4/g_*$. This means that the typical value for the inflaton field
at the end of the Planck era is $\phi \sim \alpha^{-1}T \ll g^{-1} T $
for small couplings and a sufficiently large number of species such that $\alpha \lesssim 1$. Hence, assuming an
initial equipartition of energy, the fermion mass is initially below
the temperature, and as we have shown earlier, it will then remain so
throughout the radiation-dominated period.

We will start by assuming that equipartition holds, such that our
thermalization and localization mechanism begins immediately after the
Planck era. At the end of this section, we will nevertheless consider
the alternative possibility that the inflaton carries a larger
fraction of the energy balance as the Universe emerges from this
putative quantum gravity regime, in which case the fermions are
initially heavy and cannot be part of the relativistic thermal bath.

\subsection{Evolution after inflaton-radiation equality, thermal inflation and nonlinear effects}

We have shown that, in the radiation-dominated era prior to the onset
of inflation, field modes can get localized close to the origin
(plateau) through the combined effects of the induced thermal mass and
fluctuation-dissipation terms in the equation of motion. We have
concluded that the classical field component has a very small mean
variance, $\langle \phi_c^2\rangle\simeq (6\pi^2\alpha^2)^{-1}H^3/T
\ll T$, while the full field variance including sub-horizon momentum
modes with $p\lesssim T$ is larger, $\langle \phi^2\rangle \sim
(2\pi^2\alpha^2)^{-1} T^2 \sim T^2$. The energy in the sub-horizon
modes redshifts as radiation, so it will not come to dominate over the
constant plateau $V_0$ as long as the super-horizon classical field
modes can be kept on top of the plateau for sufficiently
long. Nevertheless, since there are nonlinear terms in the scalar
potential, one must check whether the consequent mixing between sub-
and super-horizon field modes can bring the latter outside the plateau
and spoil the initial conditions for inflation.

{}First, let us note that the inflaton-radiation equality temperature
$T_{eq}$ satisfies
\begin{equation}
T_{eq} = (30/\pi^2)^{1/4} g_*^{-1/4} V_0^{1/4} \ll v,
\end{equation}
since $V_0=\lambda v^4$ with $\lambda \ll 1$. This means the full
inflaton field should be everywhere within the plateau region at
equality, also giving $g\phi \ll T$, such that fermions are light as
assumed. 

The full effective potential at finite temperature can be written as:
\begin{equation} \label{NR_thermal}
V_{eff}= V_0\left[ 1-3 (\phi/v)^4+ 2(\phi/v)^6 \right] - {2N_F\over \pi^2}T^4
\int_0^{\infty} dy
y^2\log\left[1+\exp\left(-\sqrt{y^2+g^2\phi^2/T^2}\right)\right].
\end{equation}
{}For relativistic fermions, $g|\phi| \ll T$, when including the leading thermal
corrections, we obtain that
\begin{equation} \label{NR_thermal_high}
V_{eff} \simeq V_0\left[ 1-3 (\phi/v)^4+ 2(\phi/v)^6 \right] -{7\pi^2\over 180}
N_FT^4+ {1\over 2}\alpha^2 T^2 \phi^2,
\end{equation}
while thermal corrections are Boltzmann-suppressed for $g|\phi|\gtrsim
T$. Note that the first term in this high temperature expansion is
field independent and has a negative sign. This does not, however,
correspond to a negative energy density, since the (free) energy density is
$\rho=V_{eff}+Ts$, where the entropy density is $s=-d V_{eff}/d T$. This yields a
total field independent contribution of $(7\pi^2/60)N_FT^4$, which is
the well-known contribution of $N_F$ relativistic Dirac fermions to
the energy density of the thermal radiation bath, usually written as
$\rho_R= (\pi^2/30) g_*T^4$ with $g_*=(7/8)\times 4N_F$.

\begin{figure}[htbp]
\centering\includegraphics[scale=1.1]{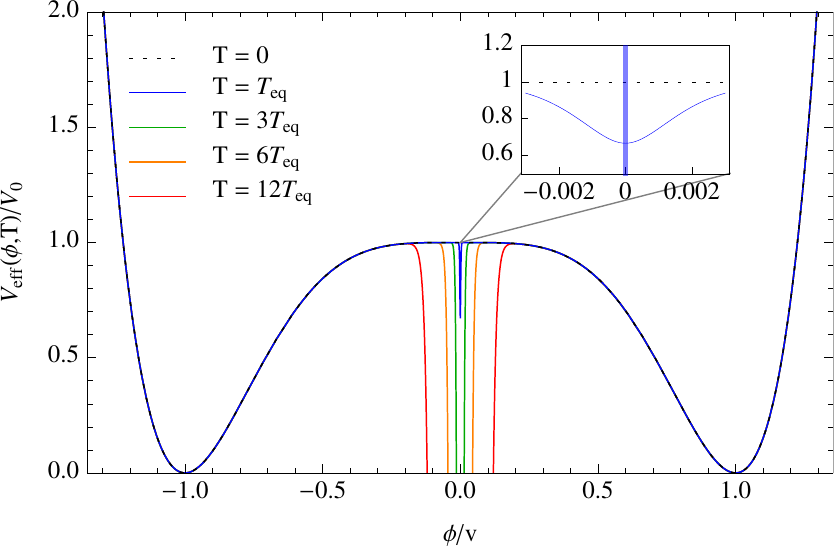}
\caption{The effective plateau-like potential given by
  Eq.~(\ref{NR_thermal}) for different temperatures above $T_{eq}$.
  The inset plot zooms in at the region about the origin for
  $T=T_{eq}$, with the narrow shaded blue region corresponding to
  $-T_{eq}<\phi<T_{eq}$ and hence to the average fluctuation range of
  the inflaton field at this temperature.}
\label{NR_potential_thermal}
\end{figure}

Let us then study the change in the shape of the potential close to
the equality temperature $T_{eq}$, which we illustrate in
{}Fig.~\ref{NR_potential_thermal} (note that when plotting the effective potential,
since $\phi$ covers values ranging from high temperature $g|\phi| \ll T$ to
low temperatures $g|\phi| \gtrsim T$, it is not appropriate to use the result
Eq.~(\ref{NR_thermal_high}), but the full expression Eq.~(\ref{NR_thermal})
or an approximation of it that covers both the high and low temperature
regimes~\cite{Borges:2016nne}).

We can see in this figure that, on the one hand, at large field values
the effective potential coincides with the zero-temperature form,
which is due to the fact that $g|\phi|\gtrsim T$ at large field values
and thermal effects are exponentially suppressed. In a vicinity of the
origin, on the other hand, there is a dip in the potential as a result
of thermal effects, essentially corresponding to the thermal mass
correction that we have considered in the pre-radiation era. This dip
naturally becomes shallower and narrower as the temperature decreases,
but it never actually disappears {\it if} the fermion mass vanishes at
the origin, since there is always an arbitrary vicinity of the origin
where the latter are relativistic and yield a thermal correction to
the otherwise vanishing inflaton mass.

Despite the thermal dip in the potential becoming increasingly narrow,
we have seen that fluctuation-dissipation effects yield an average
thermal field fluctuations $\langle |\phi|\rangle \sim T$, for which
$g|\phi|<T$ for small couplings. This means that the inflaton is
localized in this shallow dip at the equality temperature, as shown in
the zoomed inset plot in {}Fig.~\ref{NR_potential_thermal}. At such average field
values we indeed find for the first and second derivatives of $V_{eff}$
with respect to the field, respectively,
\begin{eqnarray}  \label{NR_thermalVl}
V_{eff}'(\phi\sim T, T) & \simeq & (-12\lambda + \alpha^2)T^3>0,
\\
\nonumber\\ 
V_{eff}''(\phi\sim T, T) & \simeq & (-36\lambda +\alpha^2)T^2
>0,
\label{NR_thermalVll}
\end{eqnarray}
where we have used that $\alpha^2\gg \lambda$. This implies that the
potential is dominated by the thermal mass for field values below the
temperature. Hence, at $T_{eq}$ all field modes will still oscillate
about the origin.

Since $V_0$ becomes dominant and the Hubble parameter becomes
constant, $H_{inf}\simeq \sqrt{V_0/3 M_P^2}$, the condition
$\Gamma_\phi > H_{inf}$ eventually fails and the field can no longer
remain in thermal equilibrium, such that we expect the
fluctuation-dissipation effects from the decay into fermions to become
sub-dominant. The decoupling temperature at which $\Gamma_\phi=
H_{inf}$ is then given by
\begin{eqnarray}  \label{decoupling_T}
T_D= {16\pi\over 3}{\lambda\over \alpha^4} {v^2\over M_P}\simeq 0.2
\left({\alpha\over 0.4}\right)^{-4} \left({\lambda\over
  10^{-14}}\right)^{1/4}g_*^{1/4}\left({v\over M_P}\right) T_{eq},
\end{eqnarray}
which shows that thermalization may still proceed for temperatures
parametrically below equality for $v\ll M_P$. Note that this does not
imply that all dissipative effects shut down, since adiabatic
dissipation (see, e.g., Ref.~\cite{BasteroGil:2010pb}) may still be
significant and in fact lead to a period of warm
inflation~\cite{Berera:2008ar,BasteroGil:2009ec}, as we discuss in Appendix B, but let us for
simplicity focus on the case where dissipative effects can be
neglected below $T_{eq}$. As the thermal mass dominates the potential
at $T_{eq}$ in the localized field range, all relevant momentum modes
satisfy a linear equation of motion, which is given by
\begin{eqnarray} 
 \ddot\phi_k + 3H_{inf} \dot\phi_k + \omega_k^2\phi_k=0,
\end{eqnarray}
where $\omega_k \simeq  \alpha T$ for all momentum modes with
$k\lesssim T$. As long as $T\gtrsim H_{inf}$, the modes remain
underdamped and oscillating about the origin, and it is easy to check
that the amplitude of each mode decays as $e^{-3Ht /2}$, i.e., the
whole field amplitude decays as $a^{-3/2}$. Since $T\propto 1/a$, the
field amplitude decreases faster than the temperature, and one remains
in the regime where $g|\phi | <T$ and where the thermal mass dominates
the potential until the temperature drops below $H_{inf}$. This is, in
fact, essentially a period of thermal
inflation~\cite{Lyth:1995hj,Lyth:1995ka} following the
pre-inflationary radiation-dominated era. One can also see that this
could have an arbitrarily long duration, since even though the
temperature is decreasing, the field gets more and more localized
close to the origin, where there is a local minimum (note that this is
a feature of plateau-like potentials where the zero-temperature field
mass vanishes at the origin). This period may, however, end before the
temperature falls below the Hubble parameter, if the fermions have a
field-independent mass $m_f > H_{inf}$, such that thermal effects
become exponentially suppressed for any field value for $T\lesssim
m_f$. We conclude that the average inflaton field value at $T\sim m_f$
should then be given by
\begin{eqnarray}  
\langle \phi^2\rangle \simeq T_{eq}^2\left({m_f\over T_{eq}}\right)^3,
\end{eqnarray}
where we used that $\phi \propto a^{-3/2} \propto T^{3/2}$. At this
stage, the nonlinear zero-temperature terms in the inflaton potential
become dominant and the inflaton field will classically roll towards
the true minimum at $\phi=\pm v$. To ensure that the classical
component can follow a slow-roll trajectory towards the minimum, we
may evaluate the slow-roll parameters at this average field value,
yielding in particular, that
\begin{eqnarray}  
\eta_\phi \simeq  -36\left({M_P\over v}\right)^2 \left({T_{eq}\over
  v}\right)^2\left({m_f\over T_{eq}}\right)^3,
\end{eqnarray}
where $\eta_\phi = M_P^2 V''/V$ as usual.
If one takes, e.g.,~$m_f\sim H_{inf}$, this yields
\begin{eqnarray}  
|\eta_\phi|\sim 10^{-17} g_*^{1/4} \left({v\over M_P}\right)\ll 1,
\end{eqnarray}
when taking $\lambda \sim 10^{-14}$. This implies that nonlinear
effects cannot spoil inflation along the plateau-like potential for
any value of $v< M_P$. The field is then extremely localized at the
origin after this period of thermal inflation, such that a long period
of slow-roll inflation should follow.

The duration of the period of thermal inflation is model dependent, being set by the ratio $T_{eq}/m_f$ according to the above discussion. We do not expect this period to have any direct observational impact, since curvature perturbations exiting the horizon during this period have a temperature dependent spectrum that cannot be nearly scale-invariant due to the exponentially fast decrease in the temperature. Hence, the relevant CMB perturbations must be generated during the subsequent slow-roll phase, which must last for 50-60 e-folds independently of the duration of the previous thermal inflation phase. Nevertheless, thermal inflation sets the temperature at which slow-roll begins and thus the initial state for inflaton fluctuations, which differs from the Bunch-Davies vacuum. This may then lead to distinctive observational features that we plan to analyze in detail in future work.

We note that the analysis above is purely classical, neglecting in
particular the possibility of quantum tunneling. Although this
possibility is hard to analyze in detail, we note that for tunneling
into the true minimum of the potential to occur, one requires a field
fluctuation that places the inflaton outside the range where the
thermal mass is dominant. {}For $T\gtrsim m_f$, this implies a field
fluctuation such that the fermions become heavy, $\delta\phi \sim
T/g$. {}For $g\ll 1$, such a fluctuation is much larger than the average
thermal field fluctuations, $\sqrt{\langle \phi^2\rangle }\sim T$, so
we expect the probability of quantum tunneling to be suppressed for
small couplings and parametrically large field multiplicities such
that $\alpha \sim g\sqrt{N_F} \lesssim 1$.

\subsection{An initial period of chaotic inflation for large 
field amplitudes followed by inflation on the plateau}

We note from the results of the previous subsection that the thermal mass 
is the leading correction to the
potential for $g|\phi|<T$. Other corrections in the high-temperature
expansion of the effective potential (see, e.g., Ref.~\cite{kapusta})
include, for example, Coleman-Weinberg terms at finite temperature
that induce quartic self-interactions of the inflaton field with an
effective coupling $\lambda_{eff}\sim g^4 N_F$ up to loop suppression
factors. These corrections to the quartic coupling, generated by the
loop thermal corrections to the tree-level potential, may exceed the
zero-temperature coupling $\lambda\sim 10^{-14}$ (noting that at
zero-temperature Coleman-Weinberg corrections may be suppressed by
supersymmetry or decoupling effects as discussed previously, in
Subsec.~\ref{decoupling}). These induce nonlinear effects that mix the
different field modes, but are generically suppressed compared to the
effect of the thermal mass and, hence, they may be discarded. {}For
instance, we have that
\begin{eqnarray} 
{\lambda_{eff} \phi^4 \over \alpha^2 T^2\phi^2}\sim {g^4 N_F {T^4\over
    \alpha^4}\over T^4}\sim N_F^{-1} \ll 1.
\end{eqnarray}
Although this shows that generically our analysis should be valid
already at the Planck era, we cannot exclude the possibility that the
potential is dominated by the zero-temperature power-law term
($\phi^6$ in the present case-study) at the Planck era. This may occur
in particular if, for some reason, equipartition of energy at the
Planck era does not apply and the inflaton energy density exceeds that
of the thermal bath. In this case we may have $g|\phi|\gtrsim T$ at
the Planck era, such that the potential has the zero-temperature form.

 This possibility is quite interesting, since in this case the field
 value will typically be sufficiently large to trigger a period of
 chaotic inflation with a power-law potential already at the Planck
 era. Note that $H\sim M_P$ at the Planck era, implying that only
 super-planckian modes are within the horizon at this stage. Assuming that the typical physical momentum of the excited field modes
 is at most $k\lesssim M_P$, we conclude that typical modes are
 super-horizon at this stage and so chaotic inflation will be
 triggered if the slow-roll coefficient $\epsilon_\phi =  M_P^2 (V'/V)^2/2   <1$ 
(note that chaotic inflation is an
 attractor in phase space when this condition is satisfied).  {}For a
 $\phi^6$ potential this condition corresponds to $|\phi| > 3\sqrt{2}M_P
 \sim g_*^{1/4} T$ at the Planck era, which is easily satisfied if
 $|\phi| \gtrsim T/g$ and thermal effects are sub-dominant.

If chaotic inflation begins at the Planck era, we may expect the
thermal bath to be diluted away very quickly. Nevertheless, this
inflationary period will last until $\epsilon_\phi \sim 1$. At the end
of this initial slow-roll regime, it is easy to see that $V\sim
V_0(M_P/v)^6 \gg V_0$. This large energy density will then be
transferred into radiation through the standard reheating process as
the field oscillates about the minimum(a) at $\phi=\pm v$ and decays
into fermions. If reheating is sufficiently fast (as will be the case
if the parametric resonance instability of preheating \cite{TB} is
effective), the radiation energy density will largely exceed $V_0$
after this reheating period, and the plateau potential may once more
be lifted by thermal effects.

The reheating process may be quite involved, including both
perturbative and non-perturbative effects (see, e.g., Refs.~\cite{ABCM,
  Karouby} for recent reviews), initially at nearly vanishing
temperature but then at finite temperature once the fermions that
result from the inflaton decay thermalize. Nevertheless, we may have a
good idea of the parametric regimes required to successfully reheat
the Universe and thermalize the inflaton by considering first the
$T=0$ perturbative decay of the inflaton into fermions. 

Since the model Eq.~(\ref{sextic}) is dominated by the
non-renormalizable  $\phi^6$ term after inflation,  we can assume this
form for the potential in the initial phase of oscillations. Let us
also assume that reheating will occur while the inflaton amplitude is
still sufficiently large that we may discard other terms. In this
case, the inflaton mass is $m_\phi= \sqrt{V''(\phi)}=
\sqrt{60\lambda}\phi^2/v$, while the fermion mass is
$m_f=g|\phi |$. Inflaton decay is then kinematically possible once
$m_\phi> 2m_f$, which happens for field amplitudes 
\begin{eqnarray}  \label{kinematic}
|\phi|> {1\over \sqrt{15}}{g\over \sqrt{\lambda}}v,
\end{eqnarray}
which is parametrically larger than the minimum $v$, since
$\lambda\sim 10^{-14}$ for observationally consistent models, 
unless the coupling $g$ is quite
suppressed. {}For field amplitudes larger than this value, we may
neglect the fermion masses, such that the zero-temperature
perturbative decay width is given by
\begin{eqnarray} 
\Gamma_\phi = {g^2N_F\over 8\pi} m_\phi = {3\alpha^2\over
  2\pi}(\sqrt{15\lambda}) {\phi^2\over v}.
\end{eqnarray}

{}For a field oscillating in a $\phi^n$ potential, the average kinetic
and potential energies are related by the virial theorem, $\langle
K\rangle = (n/2)\langle V\rangle$, such that in our case the average
energy density of the oscillating inflaton is
\begin{eqnarray} 
\langle \rho_\phi \rangle = 4\langle V(\phi) \rangle = 8\lambda
        {\phi^6\over v^2}.
\end{eqnarray}
Since this is the dominant component before the inflaton decays
significantly, this yields an average Hubble parameter given by
\begin{eqnarray} 
H = \sqrt{8\over 3}\sqrt{\lambda} {|\phi^3|\over v M_P}.
\end{eqnarray}
Decay will then occur when $\Gamma_\phi >H$, i.e., for field
amplitudes satisfying
\begin{eqnarray} 
|\phi| < \phi_R \equiv {9\over 4\pi}\sqrt{5\over 2} \alpha^2 M_P \sim
\alpha^2 M_P.
\end{eqnarray}
Since we require $\alpha \gtrsim 0.1$ for thermalization of the
inflaton after reheating, this is just below $M_P$ and for $v\ll M_P$
we may safely discard other terms in the potential as we assumed
above. We must, for consistency, require that the decay is kinematically possible at reheating, at $|\phi|=\phi_R$, according
to the constraint Eq.~(\ref{kinematic}), yielding the upper bound:
\begin{eqnarray} 
{v\over M_P} < {45\over 4\pi}\sqrt{3\over 2} \alpha^2
\left({\sqrt\lambda \over g}\right),
\label{voverMp}
\end{eqnarray}
which may also be regarded as an upper bound on the Yukawa coupling
$g$. Note that if this is satisfied, decay will be kinematically
possible at the end of inflation with $|\phi| \sim \sqrt{18} M_P \gg
\phi_R$.

When $\Gamma_\phi \sim H$, the inflaton will effectively decay and the
Universe becomes dominated by the inflaton's decay products. Assuming
that these decay products are light (as we will explicitly check
below)  and that they thermalize quickly through processes other than
the interactions with the inflaton field (e.g., through their
interactions with gauge particles), we can estimate the reheating
temperature $T_R$ by equating $\Gamma_\phi$ with the Hubble parameter
in a radiation-dominated Universe, which yields for $T_R$ the result
\begin{eqnarray} 
T_R = {135\over 4\pi^2} \left({3\over 8}\right)^{1/4} \alpha^3
g_*^{-1/4} \lambda^{1/4} \left({M_P\over v}\right)^{1/2} M_P
 \simeq  2\alpha^3 \left({M_P\over v}\right)^{3/2}
T_{eq},
\end{eqnarray}
which is thus parametrically larger than the equality temperature,
$T_{eq}$, at which $V_0$ dominates over the radiation energy density
for $\alpha\lesssim 1$ and $v\ll M_P$. The upper bound on $v/M_P$ in
fact yields a lower bound on the reheating temperature,  given by
$T_R\gtrsim 7 (g/10^{-6})^{3/2} T_{eq}$  \footnote{We note that non-perturbative effects, in particular a parametric resonance, may speed up the reheating process and should be taken into account in a more accurate treatment, which we leave for future work. However, backreaction effects necessarily shut down the parametric resonance once the produced particles have an energy density comparable to the inflaton's energy density, and reheating always ends in the perturbative regime. Considering only the perturbative decay of the inflaton thus yields an accurate estimate of the reheating temperature.}

Now, we must check that the fermions are indeed light at the field
amplitude and temperature at reheating. {}From the above results for
$\phi_R$ and $T_R$, we readily find that
\begin{eqnarray} 
{g\phi_R\over T_R} = {\pi\over 3\sqrt{5}}\left({2\over 3}\right)^{1/4}
\alpha^{-1} g_*^{1/4} \left({g\over \lambda^{1/4}}\right)\left({v\over
  M_P}\right)^{1/2}\lesssim 2 \sqrt{\alpha},
\end{eqnarray}
where in the last step we have used the upper bound on $v/M_P$
obtained in Eq.~(\ref{voverMp}). We thus see that the fermion mass at
reheating is at most of the order of the temperature, so that the
calculation is consistent. This also means that the thermalized
fermions will then uplift the potential with the thermal mass
correction, with the condition $g|\phi|/T$ satisfied at all subsequent
times, and localize the field at the origin, such that a second period
of inflation along the plateau may occur when the temperature
eventually falls below $T_{eq}$.

We note that the consistency of this analysis requires essentially
that the upper bound on $v/M_P$, Eq.~(\ref{voverMp}), is satisfied,
which for $\lambda\sim 10^{-14}$, can be written as
\begin{eqnarray} 
{v\over M_P} \lesssim 10^{-7}\alpha \sqrt{N_F}.
\end{eqnarray}
We note that this analysis neglects thermal corrections to the
inflaton decay width, which will be active as soon as a significant
density of thermalized fermions is achieved before reheating is
complete, and also an initial stage of preheating, which may change
these constraints. Nevertheless, we see that it is parametrically
possible to reheat the Universe after the period of chaotic inflation
so as to restore the symmetry and induce a subsequent period of
inflation along the plateau part of the potential.

The possibility of having two periods of inflation is quite
attractive, since if chaotic inflation is triggered immediately after
the Planck era any initial curvature in the Universe will be diluted
away before the second period of inflation along the plateau. In the
absence of such a period of chaotic inflation, there may be a too long
period of radiation-domination before $V_0$ becomes dominant, and if
the Universe has initially a positive curvature it may overclose
before inflation begins. This implies that some degree of tuning of
the initial curvature of the Universe is required for a successful
inflationary model with a plateau-like potential, which is naturally
less attractive.

\section{Conclusions}
\label{conclusions}

In this work we have analyzed the possibility of addressing the
initial conditions problem for inflation with a plateau-like potential
with finite temperature fluctuation-dissipation effects. We have shown
that in a pre-inflationary radiation-dominated era the coupling
between the inflaton and a thermal bath of relativistic fermions
induces (i) a thermal mass correction that dominates the potential,
(ii) a friction term corresponding to the inflaton decay into fermions
and (iii) an associated noise term that sources inflaton
fluctuations. The combined result of these effects is a localization
of inflaton modes about the origin, particularly the super-Hubble
modes, while sub-Hubble modes acquire a thermal spectrum. 

This localization effect is a direct consequence of fluctuation-dissipation dynamics driving inflaton modes towards a thermal equilibrium distribution, reproducing the results found in \cite{Albrecht:1984qt} where such a distribution was assumed. One should note that this is a non-trivial result since, while local thermal equilibrium is a valid assumption on short sub-horizon scales, nothing can {\it a priori} be said about the inflaton mode distribution on large scales. Employing the framework of fluctuation-dissipation dynamics, we have shown that super-horizon inflaton modes evolve towards an equilibrium distribution as a consequence of thermal and dissipative effects on short scales.

The formalism of fluctuation dissipation has also allowed us to conclude that a localization of the inflaton's long wavelength modes occurs provided that the coupling between the inflaton and the species in the
radiation bath is sufficiently large for thermalization to occur
before the plateau potential comes to dominate the energy balance. The
lower bound on the coupling decreases with the height of the
potential, $V_0$, since inflaton-radiation equality occurs later for
smaller $V_0$, but we have found that the effective coupling $\alpha =
g\sqrt{N_F/6}> \mathcal{O}(0.1)$ unless the tensor-to-scalar ratio is
very suppressed. Such effective couplings could, of course, be the
result of a large number of species in the thermal bath and not
necessarily of large individual couplings.

While it is commonly believed that such large couplings can spoil the
flatness of the potential and its observational predictions, we have
argued that in supersymmetric models, for example, effective couplings
of the required size do not modify the tree-level potential in a
significant way, thus consistently allowing for the localization of
the inflaton at the origin through thermal fluctuation-dissipation
effects before the onset of the slow-roll regime, independently of
whether the latter occurs in a cold or warm regime. The addition of
scalar degrees of freedom will, in fact, help the thermalization
process. We have also pointed out that, even in the absence of
supersymmetry, sufficiently large couplings may be allowed provided
that the species responsible for the inflaton thermalization become
heavy during the slow-roll regime, compared to the relevant energy
scale during inflation, which should arguably be the inflationary
Hubble scale. The quantum corrections to the scalar potential induced
by these heavy species are then suppressed compared to the
conventional Coleman-Weinberg term, in agreement with the decoupling
theorem. We note that this is naturally seen using a mass-dependent
renormalization scheme, while decoupling has to be enforced as a
separate constraint in mass-independent schemes, such as the
conventionally used minimal subtraction procedure.

We have also shown that equipartition of energy at the end of the
Planck era generically leads to initial field values such that the
fermionic species in the thermal bath are relativistic and that the
field evolution is such that they remain relativistic throughout the
radiation-dominated era and even after inflation begins. In fact, for
temperatures above the bare (i.e.~field-independent) fermion mass,
$T\gtrsim m_f$, thermal corrections dominate the inflaton potential
close to the origin and prevent slow-roll from starting. If $m_f <
T_{eq}$, we have thus found that the pre-inflationary radiation era is
followed by a period of {\it thermal inflation}. This implies that the
slow-roll regime, where the final 50-60 e-folds of inflation should
occur, starts at a much lower temperature than previously
thought. This in turn suppresses the non-linear mixing between
super-Hubble and thermalized sub-Hubble modes that could otherwise
prevent the onset of slow-roll. We have also shown that the energy
stored in the sub-Hubble modes redshifts like radiation, so that they
are effectively a part of the thermal radiation bath and become
sub-dominant at $T_{eq}$. We note that dissipative effects can still be significant during this first thermal inflation stage, until the inflaton decay rate drops below the expansion rate, and in this sense there is also an early warm inflation period even if fluctuation-dissipation effects play no role in the subsequent slow-roll dynamics.

If equipartition of energy does not hold at the Planck scale and the
inflaton carries a larger fraction of the energy density, one is
likely to start in a regime where the fermions are heavy and do not
induce thermal effects on the scalar potential. Unless the plateau
section of the potential extends to such large (and most probably
super-planckian) field values, the field will naturally be outside the
plateau region where the potential has a monomial form, as in the
examples considered in this work. It is then likely that a first
period of chaotic inflation may be triggered at the Planck era. At the
end of this period, the energy density in the inflaton still largely
exceeds the height of the sub-Planckian plateau, so that after
reheating one may enter a radiation-dominated era during which the
inflaton gets localized at the origin by fluctuation-dissipation
effects and a second period of slow-roll inflation along the plateau
may follow. CMB observations will only be sensitive to the final
period, but this is an attractive scenario since it allows for
inflation starting immediately after the Planck era while yielding a
sufficiently low tensor-to-scalar ratio.

Hence, the thermalization and localization mechanism described in this
work can be relevant whether there is already a thermalized bath at
the end of the Planck era or if inflation is triggered from chaotic
initial conditions at this stage, producing a thermal bath upon a
first reheating stage. The first possibility should occur if the
initial inflaton field value is sufficiently small for the fermions it
interacts with to be light at this stage, provided of course that
thermalization of these degrees of freedom can occur in a Planck
regime, while chaotic inflation should be triggered in patches of the
Universe where the inflaton field value is larger and the fermions are
most likely too heavy to form a thermalized bath. Both possibilities
are available in potentials with a flat section near the origin and a
monomial behavior at large field values.

While several of our conclusions were derived in the context of a
particular plateau-like potential, we expect the thermalization and
localization mechanism to apply to any potentials with a plateau-like
region about the origin such that $V'(0)=V''(0)=0$, and which require
fine-tuned initial conditions to trigger a sufficiently long period of
slow-roll inflation. Thermal fluctuation-dissipation effects yield a
natural dynamical mechanism to obtain such initial conditions and, as
we have shown in this work, can be significant without leading to
large quantum corrections to the inflaton potential, as commonly
believed. While there are potentially other aspects of the initial
condition problem at the non-linear level that we cannot address in
this framework, our conclusions show that the inflationary paradigm is
not necessarily in trouble with the latest constraints on the
tensor-to-scalar ratio, even without considering modifications of the
gravity theory. 
Furthermore if the inflationary period is just long enough to
solve the horizon problem, it has been recognized
\cite{Berera:1997wz,Bhattacharya:2005wn} that remnants 
of the pre-inflationary era might leave
imprints on the data.  The pre-inflation fluctuation-dissipation
dynamics studied in this paper could be of interest to examine in this context.
We thus hope that this work motivates further
consideration of plateau-like models in particle physics scenarios where the
generic mechanism outlined here can be implemented.


\acknowledgments    M.B.G.~is partially supported by ``Junta de
Andaluc\'ia'' (FQM101) and the University of Granada (PP2015-03).
A.B. is supported by STFC.  R.B. wishes to thank the Institute for
Theoretical Studies of the ETH Z\"urich for kind
hospitality. R.B. acknowledges financial support from Dr.  Max
R\"ossler, the Walter Haefner Foundation, the ETH Z\"urich Foundation,
and from a Simons Foundation fellowship.  The research of R.B. is also
supported in part by funds from NSERC and the Canada Research Chair
program.  I.G.M. is partially supported by  the UK Science and
Technology Facilities Council Consolidated  Grant ST/J000426/1.  R.O.R
is partially supported by research grants from Conselho Nacional de
Desenvolvimento Cient\'{\i}fico e Tecnol\'ogico (CNPq), grant
No. 303377/2013-5,  Funda\c{c}\~ao Carlos Chagas Filho de Amparo
\`a Pesquisa do Estado do Rio de Janeiro (FAPERJ), grant No. E - 26 /
201.424/2014, and Coordena\c{c}\~ao de Pessoal de N\'{\i}vel Superior
- CAPES (Processo No. 88881.119017/2016-01).  J.G.R. is supported by the FCT Investigator Grant No.~IF/01597/2015 and partially by the H2020-MSCA-RISE-2015 Grant No. StronGrHEP-690904 and by the CIDMA Project No.~UID/MAT/04106/2013. The authors also acknowledge the kind hospitality and support of the Higgs Centre for Theoretical Physics at the University of Edinburgh.

\appendix


\section{Inflation with the Higgs-like symmetry breaking potential}
\label{Sec:Higgs}

In this appendix we analyze the inflationary dynamics for a
Higgs-like potential of the form
\begin{equation}
V(\phi)={\lambda\over 4}(\phi^2-v^2)^2.
\end{equation}
{}For this potential, the slow-roll parameters are given by
\begin{equation}
\epsilon_\phi=8 \left({M_P\over v}\right)^2 {x^2\over (1-x^2)^2}, \qquad \eta_\phi=
-4 \left({M_P\over v}\right)^2 {1-3x^2\over (1-x^2)^2},
\end{equation}
where $x=\phi/v$.
From this we immediately see that the potential suffers from an
``eta-problem" and cannot sustain a long period of
inflation for  $v<2 M_P$ even if the field is initially very close to
the origin. We thus expect a
successful period of inflation to occur only for super-planckian values of $v$. This is similar to chaotic inflation models, where
the slow-roll conditions are only satisfied for super-planckian field
values, but in this case inflation may actually occur for
sub-planckian field values, the condition $v\gtrsim M_P$ being
essentially a condition on the curvature of the potential near the
origin. 

\begin{figure}[htbp]
\centering\includegraphics[scale=1.2]{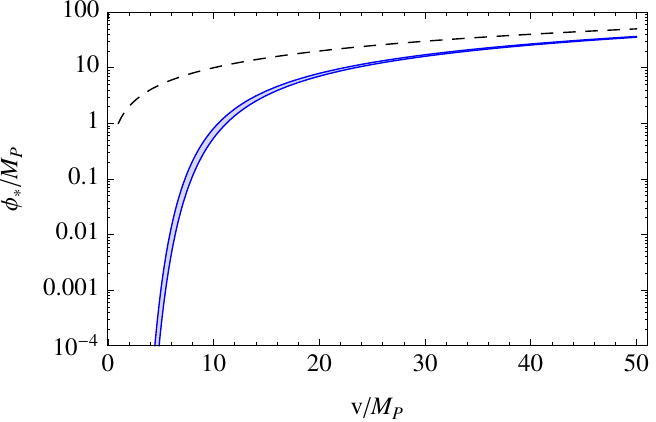}
\caption{Field value at horizon-crossing yielding 50-60 e-folds of
  inflation as a function of $v$.  {}For comparison, the dashed black
  line gives the value of $v$.}
\label{horizon_field}
\end{figure}

The number of e-folds of inflation after the present
CMB scales become super-horizon at a field value $\phi_*$ is given by
\begin{equation}
N_e\simeq -\int d\phi {V(\phi)\over V_{,\phi}(\phi)} = {v^2 \over
  {8M_P^2}}\left[\ln\left({x_e^2\over
    x_*^2}\right)+x_*^2-x_e^2\right],
\end{equation}
where the final field value $\phi_e$ corresponds to the
smallest field value for which $\epsilon_\phi=1$. In {}Fig.~\ref{horizon_field} we plot the field value at horizon-crossing
for 50-60 e-folds of inflation. As one can easily see, the last 50-60
e-folds of inflation will occur with super-planckian fields for
$v\gtrsim 10 M_P$. 

\begin{figure}[htbp]
\centering\includegraphics[scale=1.2]{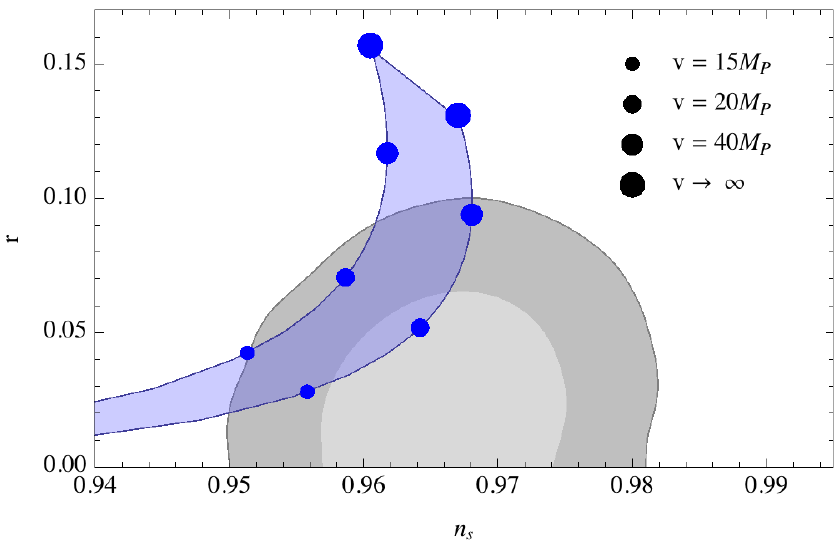}
\caption{Trajectories in the $(n_s,r)$ plane for the symmetry-breaking
  potential with $50-60$ efolds of inflation (shaded blue region),
  along with the 68\% and 95\% C.L.~obtained by the Planck
  collaboration. The circles give the prediction for particular values
  of $v$.}
\label{higgsnsr}
\end{figure}

The spectral index and tensor-to-scalar-ratio are given, respectively,
by
\begin{eqnarray}
n_s&=&1-6\epsilon_{\phi_*}+2\eta_{\phi_*} = 1- 8
\left({M_P\over v}\right)^2 {4x_*^2+ 1 \over (x_*^2-1)^2},\\ r&=&16\epsilon_{\phi_*}=48 \left({M_P\over v}\right)^2
      {x_*^2\over (x_*^2-1)^2}.
\end{eqnarray}
In  {}Fig.~\ref{higgsnsr} we show the observational predictions for $50-60$ e-folds
of inflation after horizon-crossing and the constraints obtained by the Planck
collaboration~\cite{Ade:2015lrj}. It is clear in {}Fig.~\ref{higgsnsr}
that the model is only in agreement with observations for $ 15
M_P\lesssim v \lesssim 40M_P$, which corresponds to the regime where
the observable e-folds of inflation occur for super-planckian field
values.

\begin{figure}[htbp]
\centering\includegraphics[scale=1.15]{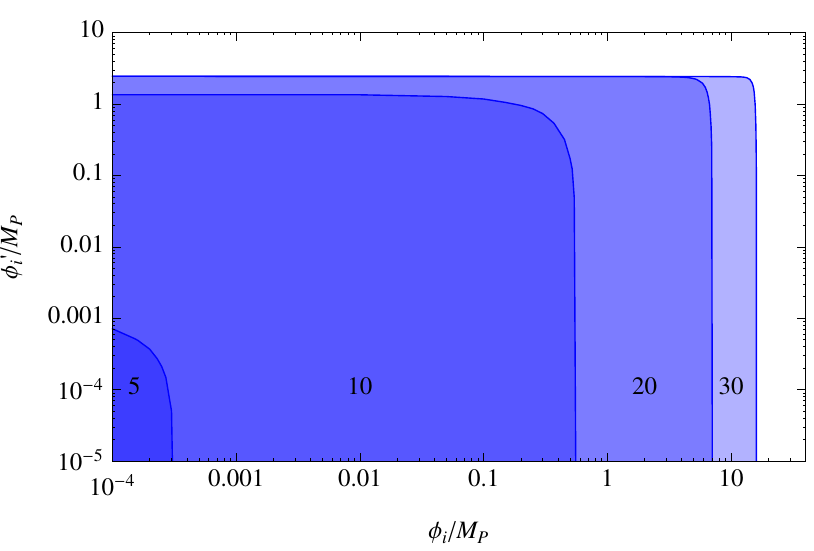}
\caption{Regions in the initial phase space plane $(\phi_i, \phi_i')$
  for which inflation lasts more than 60 e-folds for different values
  of $v/M_P$ indicated by the numbers inside each region. }
\label{higgsphidotphi}
\end{figure}

With the observational constraints on the value of the inflaton field
VEV in mind, we may now discuss the initial conditions in phase space
that lead to a sufficiently long period of inflation.
We have solved the scalar field equation,
\begin{eqnarray}
  \ddot\phi+3H\dot\phi+V_{,\phi}(\phi)=0,
  \label{field_eq}
\end{eqnarray}
for different initial conditions in phase space and for different
values of $v$. We show in {}Fig.~\ref{higgsphidotphi} the regions
in the plane $(\phi_i, \phi_i')$, with the subscript $i$ denoting the
initial value and $\phi'=\dot \phi/H$, where
the slow-roll condition $\epsilon_\phi<1$ fails after more than 60
e-folds.

As one can clearly see in  {}Fig.~\ref{higgsphidotphi}, initial
conditions leading to a sufficiently long period of inflation need to
be significantly more tuned for smaller values of $v$, where the
curvature of the potential near the origin is larger. This is the
origin of the statement often found in the literature that ``new
inflation" potentials, as the symmetry-breaking potential example under
consideration, require extremely fine-tuned initial conditions, but as
one can clearly see this applies only to small values of $v$.

\begin{center}
\begin{figure}[htb]
\subfigure[]{\includegraphics[scale=0.9]{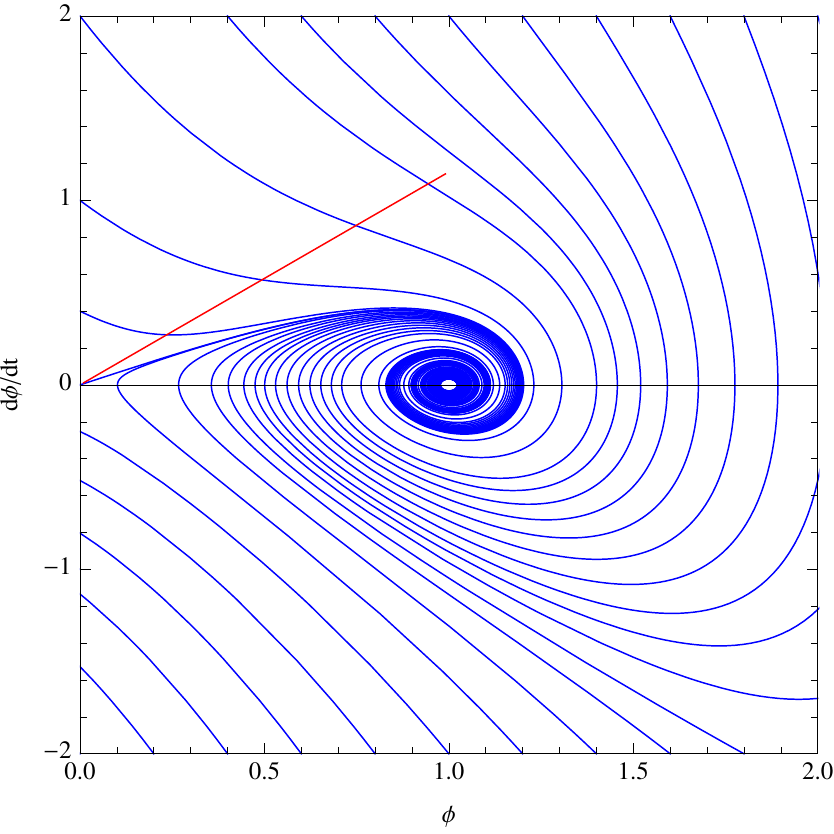}}
\subfigure[]{\includegraphics[scale=0.9]{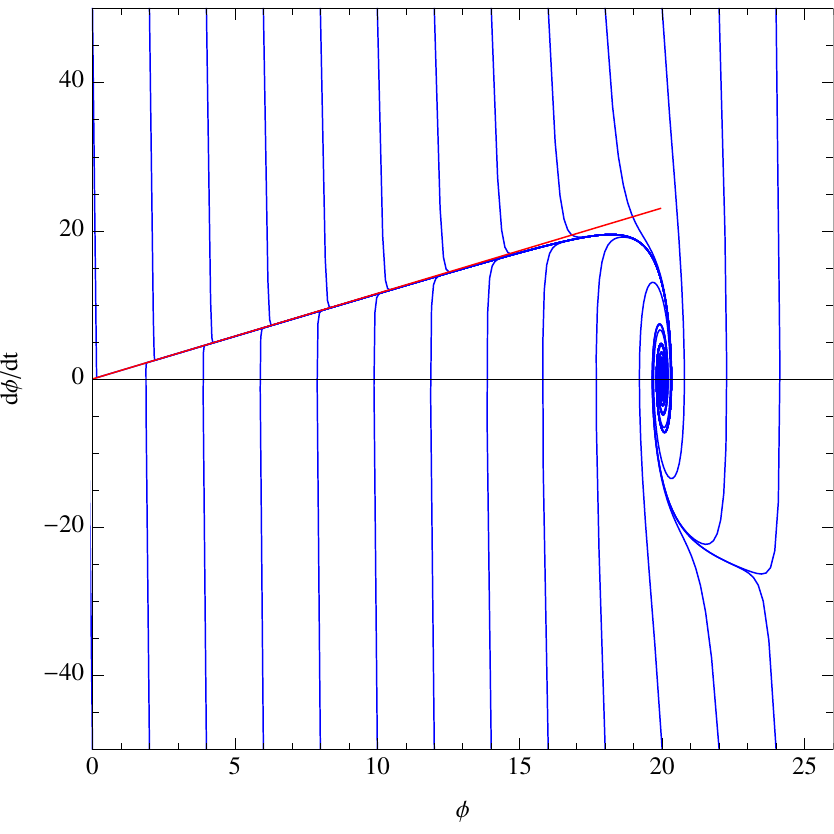}}
\caption{Phase space trajectories (shown in $M_P$ units) for $v=M_P$,
  panel (a) and for $v=20M_P$, panel (b).  Time is rescaled by a
  factor $\sqrt{\lambda}$ with respect to cosmic time. The red line
  gives, in each case, the slow-roll trajectory.}
\label{phase_space-Higgs}
\end{figure}
\end{center}

On the other hand we have seen above that observational constraints require
super-Planckian values, in which case enough inflation may occur for a
wide range of initial conditions. For
large $v$, even super-planckian field values and velocities lead to a
sufficiently long inflationary regime, with $\phi'< \sqrt{6}M_P$ for
$v\gg M_P$. In fact, in this case the
inflationary trajectory is an attractor. We show this in
{}Fig.~\ref{phase_space-Higgs}, where we plot several trajectories in
the $(\phi,\dot\phi)$ plane.  We have used a rescaled time
coordinate $\tilde{t}=\sqrt{\lambda} t$ which eliminates the
dependence of the field equation on the self-coupling $\lambda$.

As one can see in {}Fig.~\ref{phase_space-Higgs}, for $v=20M_P$, panel
(b), all trajectories with $\phi\lesssim v$ have a common straight
portion corresponding to the slow-roll solution, showing that this is
an attractor. This is opposed to the case $v=M_P$, panel (a), where no
trajectory actually follows the slow-roll solution
$\dot\phi=-V_{,\phi}/3H$. This is because for $v=M_P$ we have
$|\eta_\phi|>1$, so that there are no consistent slow-roll solutions.

We thus conclude that the statement in the literature than ``new
inflation" requires much more fine-tuning of initial conditions than
chaotic inflation is connected with the assumption that $v <
M_P$. This thus seems a wrong comparison, since chaotic inflation
requires super-planckian field values to allow for a wide range of
initial conditions that yield a sufficiently long period of
inflation. Whether the fact that the required field values are
super-planckian poses a problem is still an object of debate, in the
absence of a consistent ultra-violet completion. There are, on the one
hand, those who argue (see, e.g., Ref.~\cite{Kehagias:2014wza}) that,
since the relevant energy density and curvature of the potential are
sub-planckian, there is not any real problem in terms of putative
quantum gravity corrections to the potential. On the other hand, there
is a worry that such corrections may involve powers of $\phi/M_P$, in
which case the dynamics of inflation in the super-planckian regime may
not be under control. We do not aim to solve this problem, but we
point out that the problem is as severe for new and chaotic inflation
and that, provided that super-planckian values are allowed, both cases
may yield a sufficiently long period of accelerated expansion for a
wide range of initial conditions.


\section{Non-renormalizable plateau-like potential}
\label{plateauV}

The example of the symmetry breaking potential is, as we have seen
above, only observationally consistent for $v> M_P$, for which the
slow-roll solution is an attractor for a wide range of initial field
and velocity values. In this sense, the initial condition problem is
not very severe in this case. Let us now explore a different example,
based on a non-renormalizable potential given by
\begin{eqnarray} \label{NR_plateau_general}
  V(\phi)=V_0\left(1-{m\over m-n}x^n+{n\over m-n}x^m\right),
  \qquad x=\phi/v,
\end{eqnarray}
for $m>n$ and where $m>4$ and these potentials have minima at $\phi= 
\pm v$  with zero cosmological constant.  Let us first investigate the
case with $m=6$ and then the case of general (integer) power $m$. 

\subsection{Plateau-like potential with sextic power}

Considering $m=6$ and $n=4$ in Eq.~(\ref{NR_plateau_general}), we obtain
\begin{eqnarray} \label{NR_potential}
V(\phi)= V_0 (1-3x^4 +2 x^6), \qquad x=\phi/v,
\end{eqnarray}
and its shape is illustrated in {}Fig.~\ref{sextic:fig}. The
slow-roll parameters for this potential are given by
\begin{eqnarray} \label{NR_slowroll}
\epsilon_\phi = 72\left({M_P\over v}\right)^2\left[{x^3(x^2-1)\over
    1-3x^4+2x^6}\right]^2,  \qquad \eta_\phi = 12\left({M_P\over
  v}\right)^2\left[{x^2(5x^2-3)\over 1-3x^4+2x^6}\right],
\end{eqnarray}
such that both parameters vanish at the origin, making it a flatter
plateau than the Higgs-like potential. The number of e-folds of
inflation after horizon-crossing is given by
\begin{eqnarray} \label{NR_efolds}
N_e ={1\over 12}\left({v\over M_P}\right)^2\left[{1\over
    2x_*^2}-{1\over 2x_e^2}+x_*^2-x_e^2-\ln\left({x_*\over
    x_e}\right)\right],
\end{eqnarray}
where $x_e\equiv \phi_e/v$ is the real solution of $\epsilon_\phi=1$
closer to the origin (we choose positive $x$ for concreteness using
the reflection symmetry of the potential).

The spectral index and tensor-to-scalar ratio are given, respectively,
by
\begin{eqnarray} \label{NR_observables-ns-r}
n_s &=&1- 432\left({M_P\over v}\right)^2\left[{x^3(x^2-1)\over
    1-3x^4+2x^6}\right]^2+ 24\left({M_P\over
  v}\right)^2\left[{x^2(5x^2-3)\over
    1-3x^4+2x^6}\right],\nonumber\\ r&=& 1152\left({M_P\over
  v}\right)^2\left[{x^3(x^2-1)\over 1-3x^4+2x^6}\right]^2.
\end{eqnarray}

\begin{figure}[htbp]
\centering\includegraphics[scale=0.32]{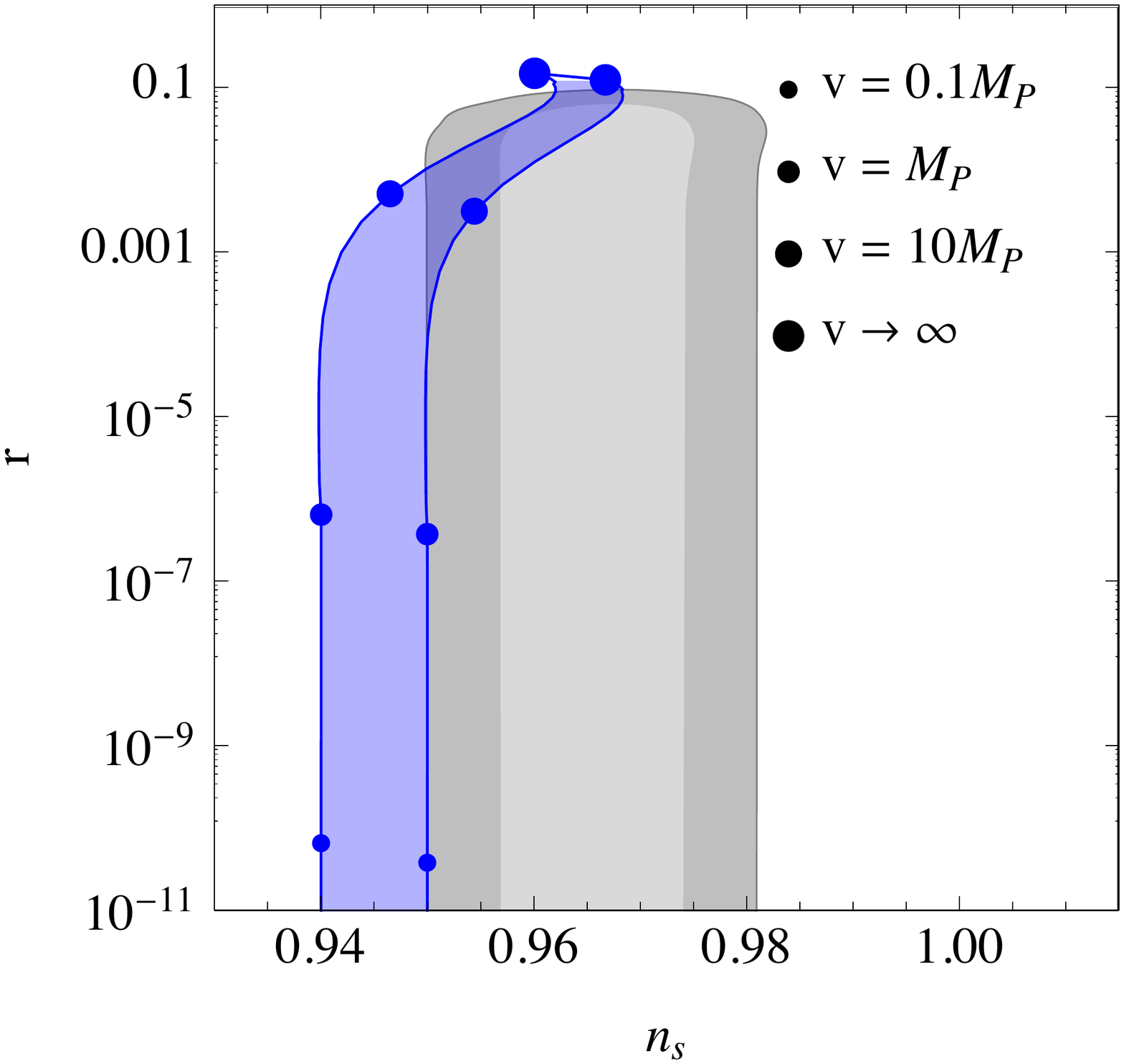}
\caption{Trajectories in the $(n_s,r)$ plane for the non-renormalizable plateau-like
  potential with $50-60$ efolds of inflation (shaded blue region),
  along with the 68\% and 95\% C.L.~obtained by the Planck
  collaboration. The circles give the prediction for particular values
  of $v$.}
\label{NR_observables}
\end{figure}

It is then easy to see that for $v\ll M_P$ we have $x_*\ll 1$ and that
$n_s-1 \simeq 3/N_e$, which yields $n_s=0.94-0.95$ for 50-60 e-folds
of inflation and an extremely small tensor-to-scalar ratio
$r\simeq (v/M_P)^4/(12 N_e^3)\ll 1$. Although the values of the
spectral index in this sub-planckian regime are just outside the 95\%
C.L. Planck contour, as shown in {}Fig.~\ref{NR_observables}, allowing
for a slightly larger number of e-folds would bring this potential
into agreement with the data.
As one can see, for $10M_P\lesssim v \lesssim 20 M_P$
there is a good agreement with the data, but this case is similar
to the Higgs-like potential analyzed above and we do not expect
 a severe initial condition problem in this regime. We will
then focus on the $v< M_P$ regime. 

\begin{figure}[h]
\centering\includegraphics[scale=1.0]{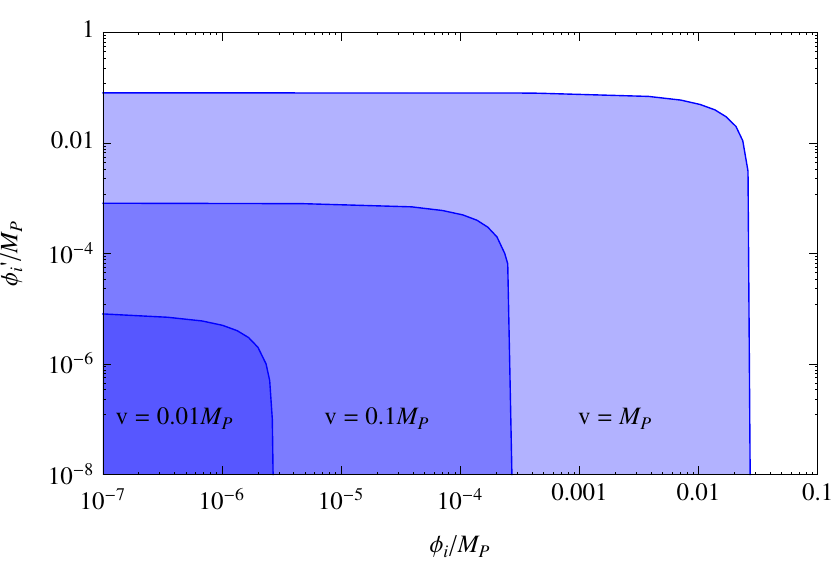}
\caption{Regions in initial phase space $(\phi_i, \phi_i')$ that lead
  to over 60 e-folds of inflation for different values of $v\leq
  M_P$.}
\label{initial_cond_NR_60}
\end{figure}

As for the Higgs-like potential, we have solved the 
(homogeneous) scalar field equation
(\ref{field_eq}) numerically for different initial conditions
in phase space, searching for values for which slow-roll lasts more than 60 e-folds. The
results are illustrated in Fig.~\ref{initial_cond_NR_60}.
It is clear that, for smaller $v$,
one requires the field to be initially closer to the origin and with a
smaller velocity.

\begin{center}
\begin{figure}[htb]
\subfigure[]{\includegraphics[scale=1.1]{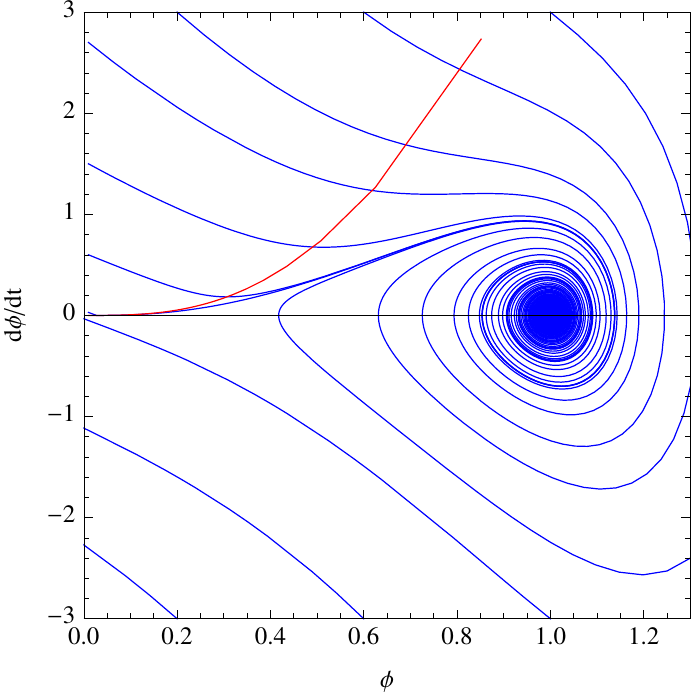}}
\subfigure[]{\includegraphics[scale=1.1]{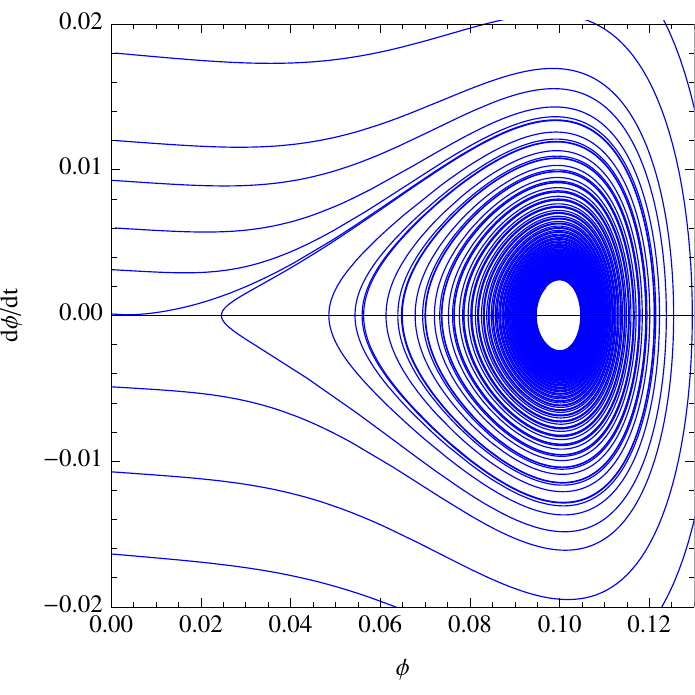}}
\caption{Phase space trajectories (shown in $M_P$ units) for the
  non-renormalizable plateau-like potential with  $v=M_P$, panel (a)
  and for $v=0.1 M_P$, panel (b).  Time is rescaled by a factor
  $\sqrt{\lambda}$ with respect to cosmic time. The red line gives, in
  each case,  the slow-roll trajectory.}
\label{phase_space-sextic}
\end{figure}
\end{center}

We show in {}Fig.~\ref{phase_space-sextic} examples of phase space
trajectories for the non-renormalizable potential with two different
values of $v$. We use a re-scaled cosmic time
$\tilde{t}=\sqrt{\lambda}t$, where $\lambda = V_0/v^4$. 

Figure~\ref{phase_space-sextic} clearly shows that the slow-roll
solution is not a (global) attractor for $v<M_P$, but that it is
nevertheless followed for sufficiently small values of the field and
velocity as obtained above. Note that for the example with $v=0.1M_P$,
panel (b) in {}Fig.~\ref{phase_space-sextic}, there are trajectories
that oscillate between the two minima before settling into one of
them. This, of course, also occurs for any $v=M_P$ (and other values)
but only for larger values of the velocity than shown in this
example.


\subsection{General non-renormalizable plateau-like potentials}

\begin{figure}[htbp]
\centering\includegraphics[scale=1.1]{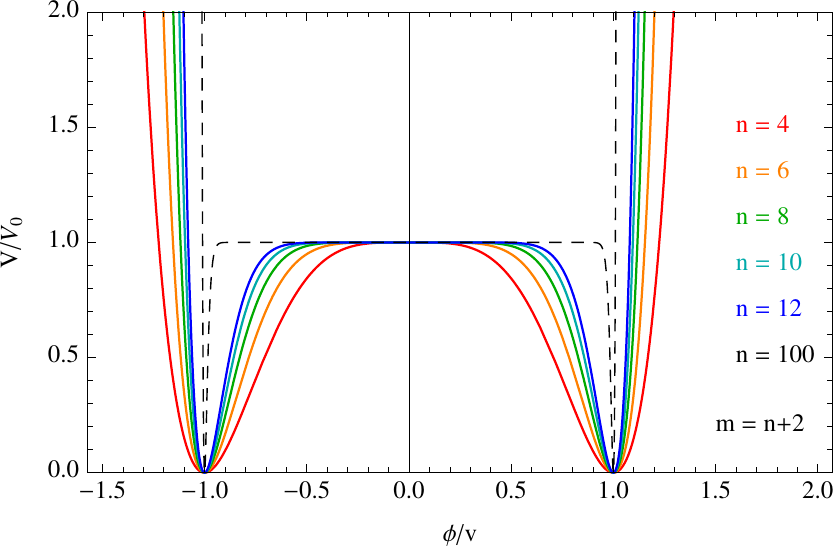}
\caption{Non-renormalizable plateau-like potentials with different
  leading powers $n$ for $m=n+2$.}
\label{plateaugeneral}
\end{figure}

Complementing the result of the previous subsection, we may consider a
generic class of non-renormalizable plateau-like potentials of the
form of Eq.~(\ref{NR_plateau_general}) with arbitrary (integer) value
$m\geq 6$.  As illustrated in {}Fig.~\ref{plateaugeneral}, for the
case $m=n+2$, these potentials yield wider plateaux the larger the
power $n$. In particular, the plateau region extends all the way to
the minima in the limit $n\rightarrow +\infty$.

{}For such potentials, we have the slow-roll parameters
\begin{eqnarray} \label{NR_plateau_general_slowroll}
\epsilon_\phi&=& {1\over2}\left({M_P\over v}\right)^2 \left({mn\over
  m-n}\right)^2 x^{2n-2}\left({x^{m-n}-1\over 1-{m\over
    m-n}x^n+{n\over m-n}x^m}\right)^2, \nonumber\\ \eta_\phi&=&
\left({M_P\over v}\right)^2 \left({mn\over m-n}\right)
x^{n-2}\left[{1-n + (m-1)x^{m-n}-1\over 1-{m\over m-n}x^n+{n\over
      m-n}x^m}\right].
\end{eqnarray}
The number of e-folds of inflation can be written in closed form as
\begin{eqnarray} \label{NR_plateau_general_efolds}
N_e&=&-{1\over M_P^2}\int_{\phi_*}^{\phi_e}d\phi{V(\phi)\over
  V'(\phi)}\nonumber \\
&=&  -{v^2\over M_P^2}{1\over mn}
\left\{{m\over
  2}x^2-\mathrm{B}\left[x^{n-m},{(m-2)\over(m-n)},0\right]
+\mathrm{B}\left[x^{n-m},-{2\over(m-n)},0\right]\right\}
\Bigr|_{x_*}^{x_e},
\end{eqnarray}
where $\mathrm{B}(z,a,b)=\int_{0}^z t^{a-1}(1-t)^{b-1} dt$ is the
incomplete beta-function~\cite{stegun}  and $x_e\equiv \phi_e/v <1$
yields $\epsilon_\phi=1$.

{}For the case $v\lesssim M_P$, we have $x_*\ll x_e <1$, and we may
write the relevant observables and number of e-folds after
horizon-crossing approximately as
\begin{eqnarray} \label{NR_observables_efolds}
n_s &=& 1-6\epsilon_{\phi_*}+2\eta_{\phi_* }\simeq1
-2(n-1)\left({mn\over m-n} \right)\left({M_P\over v}\right)^2
x_*^{n-2},\nonumber\\ r&=&16\epsilon_{\phi_*} \simeq 8  \left({mn\over
  m-n} \right)^2\left({M_P\over v}\right)^2
x_*^{2n-2},\nonumber\\ N_e&\simeq&  \left({m-n\over mn}
\right)\left({v\over M_P}\right)^2{1\over n-2} x_*^{2-n},
\end{eqnarray}
such that we may write
\begin{eqnarray} \label{NR_observables_efolds_1}
n_s &\simeq & 1-{2\over N_e}\left({n-1\over n-2}\right),
\nonumber\\ r&\simeq & 8  \left({mn\over m-n}  \right)^{-{2\over
    n-2}}\left({v\over M_P}\right)^{2n\over n-2}{1\over
  \left[(n-2)N_e\right]^{2n-2\over n-2}}.
\end{eqnarray}
We see that the scalar spectral index is independent of the power $m$,
while the tensor-to-scalar ratio exhibits a mild dependence on this
exponent. {}For $n\gg 1$, we have
\begin{eqnarray} \label{NR_observables_efolds_2}
n_s \simeq  1-{2\over N_e}, \qquad r\simeq  {8\over n^2} \left({v\over
  M_P}\right)^{2}{1\over N_e^2}.
\end{eqnarray}
We thus see that the large-$n$ prediction of this general class of
potentials for the scalar spectral index coincides with that of the
Starobinsky model \cite{Guth}, with $n_s=0.96-0.967$ for $50-60$ e-folds of
inflation, which is in very good agreement with the Planck
results. The tensor-to-scalar ratio is, in general, very suppressed
for these potentials.

\begin{figure}[htbp]
\centering\includegraphics[scale=1.1]{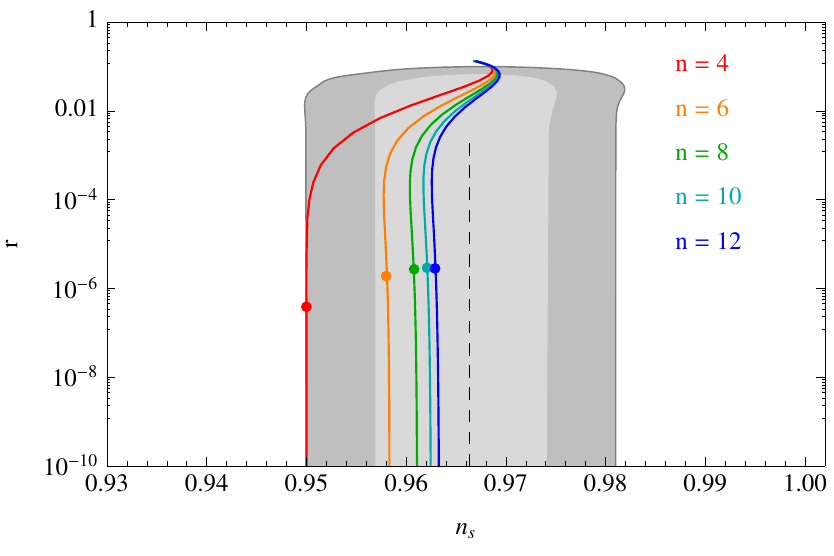}
\caption{Observational predictions for non-renormalizable plateau-like
  potentials with different leading powers $n$ for $m=n+2$ and
  $N_e=60$. The ratio $v/M_P$ decreases from top to bottom in each
  curve. The circles give, in each case, the prediction for $v=M_P$ as
  a reference. The dashed black line gives the limiting prediction for
  $n_s$ with $v\lesssim M_P$.}
\label{NR_observables_general}
\end{figure}

In {}Fig.~\ref{NR_observables_general}, we illustrate the
predictions obtained for general values of $v/M_P$ with different
powers $n$ for $m=n+2$ and $N_e=60$, where the approach to the
Starobinsky value of $n_s$ is clear for $v\ll M_P$. We also see that
the predictions for $v\gg M_P$ converge for the different values of
$n$,  and although the limiting value of $n_s$ is observationally
consistent,  the corresponding tensor-to-scalar ratio $r\simeq 0.13$
is already excluded. 

This shows that, even though the quartic plateau potential is only
marginally consistent with observational data, there exist similar
potentials that are in perfect agreement with Planck in a
sub-planckian regime. These potentials exhibit, moreover, a wider
plateau and hence are less constraining in terms of initial conditions
that can lead to a sufficiently long inflationary period. Our earlier
analysis of initial conditions for the quartic plateau provides, thus,
a worst-case scenario. Such non-renormalizable potentials should be,
for example, ubiquitous in supersymmetric theories, where are there
are several flat directions in scalar field space at the
renormalizable level and which are only lifted by non-renormalizable
terms.

\subsection{Non-renormalizable plateau model in warm inflation}

{}For completeness, we analyze the inflationary dynamics with the
non-renormalizable plateau potential (\ref{NR_potential})
in the case where dissipative effects cannot be neglected during the
slow-roll phase. {}For a slowly rolling field, the main dissipative
effects do not correspond to the field decay width as we have used for
the oscillating field modes in the pre-inflationary radiation era, but
arise from adiabatic dissipation that is non-vanishing at finite temperature~\cite{Berera:2008ar}.  In
fact, such dissipative effects transfer a part of the inflaton's
(kinetic) energy into the radiation bath, preventing it from being
diluted. Hence, if there is sufficient dissipation in this adiabatic
regime, a thermal bath can be sustained during inflation, leading to
what is known as a warm rather than cold inflationary scenario.

The presence of the thermal bath must not, however, significantly
alter the form of the inflaton potential during the slow-roll phase,
and in particular the thermal mass that drove the evolution of the
inflaton modes before the onset of inflation must become suppressed in
the slow-roll regime. This may be possible if the fields coupled to
the inflaton become non-relativistic in the slow-roll phase, with
$T\lesssim g\phi$ neglecting bare masses. Successful models of warm
inflation in this low-temperature regime have been constructed in the
context of supersymmetric models, leading to Yukawa interactions of
the form we have considered in the pre-inflationary era and their
scalar counter-parts. We have seen that supersymmetry may be indeed a
way of controlling the flatness of the inflaton potential at
zero-temperature and of keeping the inflaton self-coupling
sufficiently small while allowing for significant interactions with
other fields. The same is true at finite temperature in the
low-temperature regime.

The superpotential yielding the relevant interactions is of the form
\begin{eqnarray}  
W= \sum_{i,j}\left[g\Phi X_i^2 + hX_iY_j^2\right],
\end{eqnarray}
where the inflaton is the scalar component of the $\Phi$ chiral
superfield and the $N_X$ fermions and their scalar superpartners form the $X_i$ chiral superfields. The second term in the
superpotential containing the $Y_j$ chiral fields allows for the decay
of the $X_i$ fields. The mass of the $Y_j$ fields in the thermal bath does not depend on the inflaton field value and they can, thus, be kept light, while the $X_i$
fields become heavy in the slow-roll phase. Note that in the pre-inflationary era the inflaton is close to the origin and the $X_i$ fields are light, as assumed in our earlier analysis.

It has been shown that in the low-temperature regime the scalar
components give the largest contribution to the adiabatic dissipation
coefficient (namely from virtual modes), yielding a dissipation
coefficient of the form~\cite{Berera:2008ar,BasteroGil:2012cm}
\begin{eqnarray}  
\Upsilon= C_\phi {T^3\over \phi^2}, \qquad C_\phi =0.04 h^2 N_Y N_X\,.
\end{eqnarray}
We note that the corrections to the inflaton potential are determined
by the t'Hooft coupling $g^2N_X\lesssim 1$ and that one also requires
$h^2N_Y\lesssim 1$ for consistency of the perturbative calculation of
the dissipation coefficient. Nevertheless, the constant $C_\phi$ can
be made arbitrarily large by taking a large number of $X$ fields
weakly coupled to the inflaton, i.e.~$N_X\gg 1$, $g\ll 1$ while
keeping $\alpha^2 \sim g^2 N_X$ with not too suppressed values as we
have obtained earlier for thermalization of the inflaton field modes
prior to inflation.

The dynamics of warm inflation with this dissipation coefficient is
determined by the coupled system of inflaton and radiation equations
along with the Friedmann equation, which in the slow-roll regime can
be written as 
\begin{eqnarray}
3H(1+Q)\dot\phi\simeq -V'(\phi), \qquad \rho_R\simeq
{3\over4}Q\dot\phi^2, \qquad H^2\simeq {V(\phi)\over 3M_P^2},
\label{slowrollwarm}
\end{eqnarray} 
where $Q=\Upsilon/3H$. Consistency of the
approximation requires in this case the slow-roll conditions
$\epsilon_\phi$, $|\eta_\phi| <1+Q$ to be satisfied.
One can derive from these equations the evolution equation for the
dissipative ratio $Q$:
\begin{eqnarray}
{Q'\over Q}={1\over 1+7Q}\left(10\epsilon_\phi -6\eta_\phi
+8\sigma_\phi\right),
\end{eqnarray} 
where prime denote e-fold derivative
and we have defined the additional slow-roll parameter
$\sigma_\phi = M_P^2 (V'/ V\phi)< 1+Q$.

For the non-renormalizable plateau potential with sextic power, 
the slow-roll parameters are given in Eq.~(\ref{NR_slowroll}), which 
together with
\begin{equation}
  \sigma_\phi \simeq -12\left({M_P\over v}\right)^2 \frac{x^2
    (1-x^2)}{1-3 x^4 + 2 x^6} \,,
\end{equation}
allows us to write the slow-roll equation for $Q$ as 
\begin{eqnarray}
  {Q'\over Q}\simeq {120\over 1+7Q} \left({M_P\over v}\right)^2 x^2
  \label{dQNe}
\end{eqnarray} 
for $x=\phi/v \ll 1$. Therefore $Q^\prime>0$ and the dissipative ratio
grows during inflation. Similarly, using the form of the dissipation
coefficient and the slow-roll equation for $Q$, one obtains for the
ratio $T/H$: 
\begin{eqnarray}
{d \ln( T/H)\over dN_e} &=& {2\over 1+7Q}\left[{2+4Q\over
    1+Q}\epsilon_\phi -\eta_\phi + {1-Q\over
    1+Q}\sigma_\phi\right] \simeq  {48\over 1+7Q}
\left({M_P\over v}\right)^2 x^2 \left({1+2Q\over 1+Q}\right)~,\nonumber\\
\label{dTNe}
\end{eqnarray} 
such that also the ratio $T/H$ grows during warm inflation for the
non-renormalizable plateau potential, implying that inflation will
occur in the warm regime if $T_*\gtrsim H_*$. 

From the slow-roll Eqs. (\ref{slowrollwarm}) one can derive a relation
between the inflation field and $Q$ \cite{BasteroGil:2009ec}, and with
the help of that to integrate Eq.~(\ref{dQNe}) to obtain the number of
e-folds.  Since $Q'>0$, we may consider the case
where $Q_*\ll 1 $ and $Q_e \gg 1$, which gives to leading order 
\begin{eqnarray}
N_e\simeq {1\over 24}\left({v\over M_P}\right)^2 x_*^{-2}\left(1+ 2.8 
Q_*^{1/5}\right). 
\end{eqnarray} 
The case $Q_*\ll 1$ is particularly relevant for computing the
spectrum of curvature perturbations generated during inflation, since
in this regime one may neglect the coupling between inflaton and
radiation fluctuations. In the warm regime $T_*\gtrsim H_*$, for
thermal inflaton fluctuations the dimensionless power spectrum has
been shown to take the form~
\cite{Berera:1999ws, Hall:2003zp, Graham:2009bf,Ramos:2013nsa,Bastero-Gil:2014jsa} 
\begin{eqnarray}
\Delta_\mathcal{R}^2 = {1\over 12\pi^2}\epsilon_{\phi_*}^{-1}{V_*\over
  M_P^4} \left({T_*\over H_*}\right),
\end{eqnarray} 
which differs from the standard cold scenario by a factor $2 T_*/H_*$,
which therefore enhances the curvature power spectrum. To obtain the
scalar spectral index one must differentiate as usual the above expression with
respect to the number of e-folds of inflation (equivalent to the
logarithmic differentiation with respect to the comoving wave number
$k$). Using Eqs. (\ref{dQNe}) and (\ref{dTNe}), we have  for the
scalar spectral index
\begin{eqnarray}
  n_s -1 & =&  2\sigma_{\phi_*}-2 \epsilon_{\phi_*}  \simeq  -24 \left({M_P\over v}\right)^2 x_*^2  \simeq -{1\over
    N_e}\left(1+ 2.8 Q_*^{1/5}\right),
  \label{ns}
\end{eqnarray} 
which is red-tilted. In fact, for 50-60 e-folds of inflation one
obtains $n_s\simeq 0.96-0.97$ for $Q_*\simeq 10^{-4}-10^{-2}\ll 1$ as
assumed, showing that dissipative effects during inflation may bring
this model into agreement with the Planck data even if they are still
small at the time when the relevant CMB scales cross the horizon.

Since tensor modes are unaffected by the thermal bath
during inflation, we have for the tensor-to-scalar ratio %
\begin{eqnarray}
  r&=& {8\epsilon_{\phi_*}\over T_*/H_*}
\simeq  3\times 10^{-6}
\left({1-n_s\over 0.04}\right)^3 \left({v\over M_P}\right)^4
\left({T_*\over H_*}\right)^{-1},
\end{eqnarray} 
which is extremely small, even smaller than in the corresponding
supercooled inflationary scenario. Notice that this yields a modified
consistency relation for warm inflation $r= 4 |n_t|/ (T_*/H_*)$ \cite{Bartrum:2013fia}, but
the smallness of the tensor-to-scalar ratio in this case implies that
this may be very difficult to measure in the near future.

Let us check the consistency of the approximations made in the above
calculations.  First, we assumed that $Q_e\gg 1$ at the end of
inflation. From $|\eta_\phi|=1+Q_e \simeq Q_e$ and the relation
between $\phi$ and $Q$ we obtain that
\begin{eqnarray}
Q_e \simeq
36\left({0.04\over 1-n_s}\right)^{5/2}\left({Q_*\over
  10^{-3}}\right)^{1/2},
\end{eqnarray} 
such that as assumed inflation ends in the strong dissipation regime. Secondly, we have assumed $x \equiv\phi/v \ll 1$ for the entire
duration of inflation. From Eq.~(\ref{ns}) and $|\eta_\phi| \simeq
Q_e$ we have, respectively,
\begin{eqnarray}
x_*&\simeq & \left(1-n_s\over 24\right)^{1/2} \left({v\over
  M_P}\right)\simeq 0.04 \left({v\over M_P}\right) \,,\;\;\; 
x_e \simeq  \left({0.04\over
  1-n_s}\right)^{5/4} \left({Q_*\over 10^{-3}}\right)^{1/4}
\left({v\over M_P}\right)~,
\end{eqnarray} 
such that this approximation is valid for $v<M_P$.

Finally, we assumed that the fields coupled to the inflaton are
non-relativistic, $g\phi<
T$, so as to suppress thermal corrections to the
inflaton mass in the slow-roll regime. Using the slow-roll equations, one can show that 
\begin{eqnarray}
{d \log( \phi/T)\over dN_e}  &\simeq & -{12\over 1+7Q}
\left({M_P\over v}\right)^2 x^2 \left({3+Q\over 1+Q}\right)<0,
\end{eqnarray} 
such that $g\phi/T$ decreases during inflation. At  the end of the slow-roll regime, we find:
\begin{eqnarray}
{\phi_e\over T_e}\simeq 7435 \left({g_*\over 100}\right)^{1/6} \left({
  0.04\over 1-n_s}\right)^{3/4}\left({Q_*\over
  10^{-3}}\right)^{1/12},
\end{eqnarray} 
such that we require couplings $g\gtrsim 10^{-4}$ to keep $g\phi\gtrsim T$ throughout inflation. This is
compatible with a number of species $N_X\sim 10^6-10^7$ while keeping
$g^2N_X\lesssim 1$ to avoid large quantum corrections to the
inflaton potential.



\begin{thebibliography}{99}

\bibitem{Guth}
R.~Brout, F.~Englert and E.~Gunzig,
\emph{The Causal Universe},
Gen.\ Rel.\ Grav.\  {\bf 10}, 1 (1979);\\
R.~Brout, F.~Englert and E.~Gunzig,
 \emph{The Creation Of The Universe As A Quantum Phenomenon},
Annals Phys.\  {\bf 115}, 78 (1978);\\
A.~A.~Starobinsky,
 \emph{A New Type Of Isotropic Cosmological Models Without Singularity},
Phys.\ Lett.\ B {\bf 91}, 99 (1980);\\
D.~Kazanas,
 \emph{Dynamics of the Universe and Spontaneous Symmetry Breaking},
Astrophys.\ J.\  {\bf 241}, L59 (1980);\\
 A.~H.~Guth, 
\emph{The Inflationary Universe: A Possible Solution To The Horizon And Flatness Problems},
Phys.\ Rev.\  D {\bf 23}, 347 (1981);\\
K.~Sato,
 \emph{First Order Phase Transition Of A Vacuum And Expansion Of The Universe},
Mon.\ Not.\ Roy.\ Astron.\ Soc.\  {\bf 195}, 467 (1981);\\
L.~Z.~Fang,
 \emph{Entropy Generation in the Early Universe by Dissipative Processes 
Near the Higgs' Phase Transitions},
Phys.\ Lett.\ B {\bf 95}, 154 (1980).

  

\bibitem{Mukh}
V. Mukhanov and G. Chibisov, 
\emph{Quantum Fluctuation And Nonsingular Universe}. (In Russian), 
JETP Lett.\  {\bf 33}, 532 (1981) [Pisma Zh.\ Eksp.\ Teor.\ Fiz.\  {\bf 33}, 549 (1981)].

  
\bibitem{Hinshaw:2012aka} 
  G.~Hinshaw {\it et al.} [WMAP Collaboration],
\emph{Nine-Year Wilkinson Microwave Anisotropy Probe (WMAP) Observations: 
Cosmological Parameter Results},
  Astrophys.\ J.\ Suppl.\  {\bf 208}, 19 (2013)
  doi:10.1088/0067-0049/208/2/19
  [arXiv:1212.5226 [astro-ph.CO]].

\bibitem{Ade:2015lrj} 
  P.~A.~R.~Ade {\it et al.} [Planck Collaboration],
\emph{Planck 2015 results. XX. Constraints on inflation},
  Astron.\ Astrophys.\  {\bf 594}, A20 (2016)
  doi:10.1051/0004-6361/201525898
  [arXiv:1502.02114 [astro-ph.CO]].



\bibitem{Ekp}
J.~Khoury, B.~A.~Ovrut, P.~J.~Steinhardt and N.~Turok,
 \emph{The Ekpyrotic universe: Colliding branes and the origin of the hot big bang},
Phys.\ Rev.\ D {\bf 64}, 123522 (2001)
[hep-th/0103239].

  

\bibitem{SGC}
R.~H.~Brandenberger and C.~Vafa, 
\emph{Superstrings In The Early Universe}, 
Nucl.\ Phys.\ B {\bf 316}, 391 (1989);\\
A.~Nayeri, R.~H.~Brandenberger and C.~Vafa, 
\emph{Producing a scale-invariant spectrum of perturbations in a 
Hagedorn phase of string cosmology},
Phys.\ Rev.\ Lett.\  {\bf 97}, 021302 (2006)
[arXiv:hep-th/0511140];\\
  R.~H.~Brandenberger, A.~Nayeri, S.~P.~Patil and C.~Vafa,
\emph{String gas cosmology and structure formation},
Int.\ J.\ Mod.\ Phys.\ A {\bf 22}, 3621 (2007)
[hep-th/0608121].

  

\bibitem{FB}
F.~Finelli and R.~Brandenberger,
\emph{On the generation of a scale-invariant spectrum of adiabatic  
fluctuations in cosmological models with a contracting phase},
Phys.\ Rev.\  D {\bf 65}, 103522 (2002)
[arXiv:hep-th/0112249].

  

\bibitem{Wands}
D.~Wands,
\emph{Duality invariance of cosmological perturbation spectra},
Phys.\ Rev.\ D {\bf 60}, 023507 (1999)
[gr-qc/9809062].

  

\bibitem{Penrose}
R.~Penrose,
\emph{Difficulties with inflationary cosmology},
Annals N.\ Y.\ Acad.\ Sci.\  {\bf 571}, 249 (1989).



\bibitem{Gibbons}
G.~W.~Gibbons and N.~Turok,
\emph{The Measure Problem in Cosmology},
Phys.\ Rev.\ D {\bf 77}, 063516 (2008)
[hep-th/0609095];\\
S.~M.~Carroll and H.~Tam,
\emph{Unitary Evolution and Cosmological Fine-Tuning},
arXiv:1007.1417 [hep-th].

  

\bibitem{Trodden}
T.~Vachaspati and M.~Trodden,
\emph{Causality and cosmic inflation},
Phys.\ Rev.\ D {\bf 61}, 023502 (1999)
[gr-qc/9811037];\\
L.~Berezhiani and M.~Trodden,
\emph{How Likely are Constituent Quanta to Initiate Inflation?},
Phys.\ Lett.\ B {\bf 749}, 425 (2015)
[arXiv:1504.01730 [hep-th]].

  

\bibitem{Ijjas}
A.~Ijjas, P.~J.~Steinhardt and A.~Loeb,
\emph{Inflationary paradigm in trouble after Planck2013},
Phys.\ Lett.\ B {\bf 723}, 261 (2013)
[arXiv:1304.2785 [astro-ph.CO]];\\
A.~Ijjas, P.~J.~Steinhardt and A.~Loeb,
\emph{Inflationary schism after Planck2013},
Phys.\ Lett.\ B {\bf 736}, 142 (2014)
[arXiv:1402.6980 [astro-ph.CO]].

  

\bibitem{Piran}
D.~S.~Goldwirth and T.~Piran, 
\emph{Inhomogeneity and the Onset of Inflation},
Phys.\ Rev.\ Lett.\  {\bf 64}, 2852 (1990);
D.~S.~Goldwirth and T.~Piran, 
\emph{Initial conditions for inflation},
Phys.\ Rept.\  {\bf 214}, 223 (1992).



\bibitem{Albrecht:1984qt}
A.~Albrecht and R.~H.~Brandenberger,
\emph{On the Realization of New Inflation},
Phys.\ Rev.\ D {\bf 31}, 1225 (1985);
A.~Albrecht, R.~H.~Brandenberger and R.~Matzner,
\emph{Numerical Analysis of Inflation},
  Phys.\ Rev.\ D {\bf 32}, 1280 (1985);
A.~Albrecht, R.~H.~Brandenberger and R.~Matzner,
\emph{Inflation With Generalized Initial Conditions},
  Phys.\ Rev.\ D {\bf 35}, 429 (1987).

  
\bibitem{Brandenberger:1990wu} 
  R.~H.~Brandenberger and J.~H.~Kung,
\emph{Chaotic Inflation as an Attractor in Initial Condition Space},
  Phys.\ Rev.\ D {\bf 42}, 1008 (1990).

\bibitem{Feldman}
H.~A.~Feldman and R.~H.~Brandenberger,
\emph{Chaotic Inflation With Metric and Matter Perturbations},
  Phys.\ Lett.\ B {\bf 227}, 359 (1989).
R.~H.~Brandenberger and H.~A.~Feldman,
\emph{Effects of Gravitational Perturbations on the Evolution of 
Scalar Fields in the Early Universe},
  Phys.\ Lett.\ B {\bf 220}, 361 (1989).


\bibitem{RHBICrev}
R.~Brandenberger,
 \emph{Initial Conditions for Inflation - A Short Review},
  arXiv:1601.01918 [hep-th].


\bibitem{Matzner}
H.~Kurki-Suonio, R.~A.~Matzner, J.~Centrella and J.~R.~Wilson,
\emph{Inflation From Inhomogeneous Initial Data in a One-dimensional 
Back Reacting Cosmology},
  Phys.\ Rev.\ D {\bf 35}, 435 (1987)
P.~Laguna, H.~Kurki- Suonio and R.~A.~Matzner,
\emph{Inhomogeneous inflation: The Initial value problem},
  Phys.\ Rev.\ D {\bf 44}, 3077 (1991).
H.~Kurki-Suonio, P.~Laguna and R.~A.~Matzner,
\emph{Inhomogeneous inflation: Numerical evolution},
  Phys.\ Rev.\ D {\bf 48}, 3611 (1993)
  [astro-ph/9306009].


\bibitem{East}
  W.~E.~East, M.~Kleban, A.~Linde and L.~Senatore,
\emph{Beginning inflation in an inhomogeneous universe},
  JCAP {\bf 1609}, no. 09, 010 (2016)
  [arXiv:1511.05143 [hep-th]].

  
\bibitem{Lim}
K.~Clough, P.~Figueras, H.~Finkel, M.~Kunesch, E.~A.~Lim and S.~Tunyasuvunakool,
\emph{GRChombo : Numerical Relativity with Adaptive Mesh Refinement},
  Class.\ Quant.\ Grav.\  {\bf 32}, no. 24, 245011 (2015)
  [arXiv:1503.03436 [gr-qc]];\\
K.~Clough, E.~A.~Lim, B.~S.~DiNunno, W.~Fischler, R.~Flauger and S.~Paban,
\emph{Robustness of Inflation to Inhomogeneous Initial Conditions},
  arXiv:1608.04408 [hep-th].

\bibitem{Guth:2013sya} 
  A.~H.~Guth, D.~I.~Kaiser and Y.~Nomura,
\emph{Inflationary paradigm after Planck 2013},
  Phys.\ Lett.\ B {\bf 733}, 112 (2014)
[arXiv:1312.7619 [astro-ph.CO]].

\bibitem{Linde:2014nna} 
  A.~Linde,
\emph{Inflationary Cosmology after Planck 2013},
  arXiv:1402.0526 [hep-th].

\bibitem{Berera:1995wh}
  A.~Berera and L.~Z.~Fang,
\emph{Thermally induced density perturbations in the inflation era},
  Phys.\ Rev.\ Lett.\  {\bf 74}, 1912 (1995)
  [astro-ph/9501024].


\bibitem{Berera:1995ie}
  A.~Berera,
\emph{Warm inflation},
  Phys.\ Rev.\ Lett.\  {\bf 75}, 3218 (1995)
  [astro-ph/9509049].


\bibitem{Berera:1996nv}
  A.~Berera,
\emph{Thermal properties of an inflationary universe},
  Phys.\ Rev.\ D {\bf 54}, 2519 (1996)
  [hep-th/9601134].


\bibitem{Bartrum:2013fia}
  S.~Bartrum, M.~Bastero-Gil, A.~Berera, R.~Cerezo, R.~O.~Ramos and J.~G.~Rosa,
\emph{The importance of being warm (during inflation)},
  Phys.\ Lett.\ B {\bf 732}, 116 (2014)
  [arXiv:1307.5868 [hep-ph]].

\bibitem{Bastero-Gil:2016qru} 
  M.~Bastero-Gil, A.~Berera, R.~O.~Ramos and J.~G.~Rosa,
{\it Warm Little Inflaton},
  Phys.\ Rev.\ Lett.\  {\bf 117}, no. 15, 151301 (2016)
[arXiv:1604.08838 [hep-ph]].


\bibitem{Bartrum:2014fla}
  S.~Bartrum, A.~Berera and J.~G.~Rosa,
\emph{Fluctuation-dissipation dynamics of cosmological scalar fields},
  Phys.\ Rev.\ D {\bf 91}, no. 8, 083540 (2015)
  [arXiv:1412.5489 [hep-ph]].


\bibitem{BasteroGil:2011cx}
  M.~Bastero-Gil, A.~Berera, R.~O.~Ramos and J.~G.~Rosa,
\emph{Warm baryogenesis},
  Phys.\ Lett.\ B {\bf 712}, 425 (2012)
  [arXiv:1110.3971 [hep-ph]].

\bibitem{ng1}
  C.~-H.~Wu, K.~-W.~Ng, W.~Lee, D.~-S.~Lee and Y.~-Y.~Charng,
{\it Quantum noise and a low cosmic microwave background quadrupole},
  J. Cosmol. Astropart. Phys. {\bf 02}, 006  (2007).
[astro-ph/0604292].

\bibitem{ng2}   W.~Lee, K.~-W.~Ng, I-C.~Wang and C.~-H.~Wu,
{\it Trapping effects on inflation},
  Phys.\ Rev.\ D {\bf 84}, 063527  (2011).
[arXiv:1101.4493 [hep-th]].

\bibitem{Berera:2009gy} 
A.~Berera and R.~Rangarajan,
\emph{Quantum phase of inflation},
Phys.\ Rev.\ D {\bf 87},  043514 (2013)
Erratum: [Phys.\ Rev.\ D {\bf 87},  049901 (2013)]
[arXiv:0912.5148 [astro-ph.CO]].

\bibitem{Ramos:2013nsa} 
  R.~O.~Ramos and L.~A.~da Silva,
\emph{Power spectrum for inflation models with quantum and thermal noises},
  JCAP {\bf 03} (2013) 032.

\bibitem{Bastero-Gil:2014jsa}
  M.~Bastero-Gil, A.~Berera, I.~G.~Moss and R.~O.~Ramos,
\emph{Cosmological fluctuations of a random field and radiation fluid},
  JCAP {\bf 05} (2014) 004.

\bibitem{Bastero-Gil:2014raa} 
  M.~Bastero-Gil, A.~Berera, I.~G.~Moss and R.~O.~Ramos,
\emph{Theory of non-Gaussianity in warm inflation},
  JCAP {\bf 12} (2014) 008.

\bibitem{Vicente:2015hga} 
  G.~S.~Vicente, L.~A.~da Silva and R.~O.~Ramos,
\emph{Eternal inflation in a dissipative and radiation environment: Heated demise of eternity},
  Phys.\ Rev.\ D {\bf 93}, no. 6, 063509 (2016).

\bibitem{Berera:2000xz}
  A.~Berera and C.~Gordon,
\emph{Inflationary initial conditions consistent with causality},
  Phys.\ Rev.\ D {\bf 63}, 063505 (2001)
  [hep-ph/0010280].


\bibitem{otherpapers}
  E.~Calzetta and M.~Sakellariadou,
  \emph{Semiclassical effects and the onset of inflation},
  Phys.\ Rev.\ D {\bf 47}, 3184 (1993)
  [arXiv:gr-qc/9209007];
  C.~Appignani and R.~Casadio,
  \emph{A Radiation-like era before inflation},
  JCAP {\bf 0810}, 027 (2008)
  [arXiv:0808.0092 [gr-qc]];
  R.~Dong, W.~H.~Kinney and D.~Stojkovic,
  \emph{Symmetron Inflation},
  JCAP {\bf 1401}, 021 (2014)
  [arXiv:1307.4451 [astro-ph.CO]].

\bibitem{Collins:1991gh} 
  P.~D.~B.~Collins and R.~F.~Langbein,
\emph{On the thermodynamics of inflation},
  Phys.\ Rev.\ D {\bf 45}, 3429 (1992).



\bibitem{Carrasco:2015rva} 
  J.~J.~M.~Carrasco, R.~Kallosh and A.~Linde,
\emph{Cosmological Attractors and Initial Conditions for Inflation},
  Phys.\ Rev.\ D {\bf 92}, no. 6, 063519 (2015)
  [arXiv:1506.00936 [hep-th]].

\bibitem{Artymowski:2016ikw} 
  M.~Artymowski, Z.~Lalak and M.~Lewicki,
\emph{Multi-phase induced inflation in theories with non-minimal coupling to gravity},
  arXiv:1607.01803 [astro-ph.CO].


\bibitem{Dimopoulos:2016yep} 
  K.~Dimopoulos and M.~Artymowski,
\emph{Initial conditions for inflation},
  arXiv:1610.06192 [astro-ph.CO].

\bibitem{bellac} M. Le Bellac, \emph{Thermal field theory}, Cambridge
  Monographs on Mathematical Physics (Cambridge University Press,
  Cambridge, UK, 1996).  

\bibitem{Yokoyama:2004pf} 
  J.~Yokoyama,
\emph{Fate of oscillating scalar fields in the thermal bath and their cosmological implications},
  Phys.\ Rev.\ D {\bf 70}, 103511 (2004)
  [hep-ph/0406072].


\bibitem{BasteroGil:2010pb} 
  M.~Bastero-Gil, A.~Berera and R.~O.~Ramos,
 \emph{Dissipation coefficients from scalar and fermion quantum field interactions},
  JCAP {\bf 1109}, 033 (2011)
  [arXiv:1008.1929 [hep-ph]].
  
\bibitem{Berera:1999ws} 
A.~Berera,
\emph{Warm inflation at arbitrary adiabaticity: A Model, an existence proof for inflationary dynamics in quantum field theory},
Nucl.\ Phys.\ B {\bf 585}, 666 (2000)
[hep-ph/9904409].

\bibitem{Berera:2008ar} 
  A.~Berera, I.~G.~Moss and R.~O.~Ramos,
\emph{Warm Inflation and its Microphysical Basis},
  Rept.\ Prog.\ Phys.\  {\bf 72}, 026901 (2009)
  [arXiv:0808.1855 [hep-ph]].


\bibitem{brownian}N. G. van Kampen, {\it Stochastic Processes in Physics and
Chemistry}, 2nd ed. (North-Holland, Amsterdam, 1992).

\bibitem{Mukhanov:1990me}
  V.~F.~Mukhanov, H.~A.~Feldman and R.~H.~Brandenberger,
\emph{Theory of cosmological perturbations. Part 1. Classical perturbations. Part 2. Quantum theory of perturbations. Part 3. Extensions},
  Phys.\ Rept.\  {\bf 215}, 203  (1992).

\bibitem{Kung:1989xz} 
  J.~H.~Kung and R.~H.~Brandenberger,
\emph{The Initial Condition Dependence of Inflationary Universe Models},
  Phys.\ Rev.\ D {\bf 40}, 2532 (1989).
  
  
\bibitem{Brandenberger:1990xu} 
  R.~H.~Brandenberger, H.~Feldman and J.~Kung,
\emph{Initial conditions for chaotic inflation},
  Phys.\ Scripta T {\bf 36}, 64 (1991).
  

\bibitem{BasteroGil:2012cm} 
  M.~Bastero-Gil, A.~Berera, R.~O.~Ramos and J.~G.~Rosa,
\emph{General dissipation coefficient in low-temperature warm inflation},
  JCAP {\bf 1301}, 016 (2013)
  [arXiv:1207.0445 [hep-ph]].

\bibitem{Appelquist:1974tg} 
  T.~Appelquist and J.~Carazzone,
\emph{Infrared Singularities and Massive Fields},
  Phys.\ Rev.\ D {\bf 11}, 2856 (1975).
  
\bibitem{BasteroGil:2010vq} 
  M.~Bastero-Gil, A.~Berera and B.~M.~Jackson,
\emph{Power suppression from disparate mass scales in effective scalar field 
theories of inflation and quintessence},
  JCAP {\bf 1107}, 010 (2011)
  [arXiv:1003.5636 [hep-ph]].

\bibitem{Benetti:2016jhf} 
  M.~Benetti and R.~O.~Ramos,
\emph{Warm inflation dissipative effects: predictions and constraints 
from the Planck data},
  Phys.\ Rev.\ D {\bf 95}, no. 2, 023517 (2017)
  [arXiv:1610.08758 [astro-ph.CO]].

\bibitem{Borges:2016nne} 
  J.~S.~Borges and R.~O.~Ramos,
\emph{Symmetry breaking patterns of the 3-3-1 model at finite temperature},
  Eur.\ Phys.\ J.\ C {\bf 76}, no. 6, 344 (2016)
  [arXiv:1602.08165 [hep-ph]].


\bibitem{BasteroGil:2009ec} 
  M.~Bastero-Gil and A.~Berera,
\emph{Warm inflation model building},
  Int.\ J.\ Mod.\ Phys.\ A {\bf 24}, 2207 (2009)
  [arXiv:0902.0521 [hep-ph]].
 
 
\bibitem{Lyth:1995hj} 
D.~H.~Lyth and E.~D.~Stewart,
\emph{Cosmology with a TeV mass GUT Higgs},
Phys.\ Rev.\ Lett.\  {\bf 75}, 201 (1995)
[hep-ph/9502417].

\bibitem{Lyth:1995ka} 
D.~H.~Lyth and E.~D.~Stewart,
\emph{Thermal inflation and the moduli problem},
Phys.\ Rev.\ D {\bf 53}, 1784 (1996)
[hep-ph/9510204].

\bibitem{kapusta}J. I. Kapusta and C. Gale, {\it Finite-Temperature Field
  Theory: Principles and Applications}, (Cambridge University Press,
  Cambridge, England, 2006).

\bibitem{TB}
J.~H.~Traschen and R.~H.~Brandenberger,
\emph{Particle Production During Out-of-equilibrium Phase Transitions},
  Phys.\ Rev.\ D {\bf 42}, 2491 (1990).
  
\bibitem{ABCM}
R.~Allahverdi, R.~Brandenberger, F.~Y.~Cyr-Racine and A.~Mazumdar,
\emph{Reheating in Inflationary Cosmology: Theory and Applications},
Ann.\ Rev.\ Nucl.\ Part.\ Sci.\  {\bf 60}, 27 (2010)
[arXiv:1001.2600 [hep-th]].
  
\bibitem{Karouby}
 M.~A.~Amin, M.~P.~Hertzberg, D.~I.~Kaiser and J.~Karouby,
\emph{Nonperturbative Dynamics Of Reheating After Inflation: A Review},
  Int.\ J.\ Mod.\ Phys.\ D {\bf 24}, 1530003 (2014)
  [arXiv:1410.3808 [hep-ph]].

\bibitem{Berera:1997wz} 
A.~Berera, L.~Z.~Fang and G.~Hinshaw,
\emph{An Attempt to determine the largest scale of primordial density perturbations in the universe},
Phys.\ Rev.\ D {\bf 57}, 2207 (1998)
[astro-ph/9703020].


\bibitem{Bhattacharya:2005wn} 
K.~Bhattacharya, S.~Mohanty and R.~Rangarajan,
\emph{Temperature of the inflaton and duration of inflation from WMAP data},
Phys.\ Rev.\ Lett.\  {\bf 96}, 121302 (2006)
[hep-ph/0508070].

\bibitem{Kehagias:2014wza} 
  A.~Kehagias and A.~Riotto,
\emph{Remarks about the Tensor Mode Detection by the BICEP2 Collaboration and the 
Super-Planckian Excursions of the Inflaton Field},
  Phys.\ Rev.\ D {\bf 89}, no. 10, 101301 (2014)
  [arXiv:1403.4811 [astro-ph.CO]].

\bibitem{stegun}{\it Handbook of Mathematical Functions}, edited by M.
Abramowitz and I. A. Stegun (Dover, New York, 1972),
9th ed.


\bibitem{Hall:2003zp}
  L.~M.~H.~Hall, I.~G.~Moss and A.~Berera,
  \emph{Scalar perturbation spectra from warm inflation},
  Phys.\ Rev.\ D {\bf 69} (2004) 083525
  [astro-ph/0305015].

\bibitem{Graham:2009bf}
  C.~Graham and I.~G.~Moss,
\emph{Density fluctuations from warm inflation}, 
  JCAP {\bf 0907} (2009) 013
  [arXiv:0905.3500 [astro-ph.CO]].


\end{thebibliography}
\end{document}